\documentclass[aps,pra,showpacs,twoside,twocolumn,10pt]{revtex4-2}
\usepackage[colorlinks=true, citecolor=blue, urlcolor=blue, linkcolor = blue ]{hyperref}

\usepackage{graphicx,bm,amsmath,hyperref,xcolor,braket,plain,color,amsthm,amsfonts,float,tabularx,graphicx,bbm,mathtools,esvect,wrapfig,verbatim,enumitem,dcolumn,amssymb,appendix,physics,fmtcount,booktabs,csquotes,epsfig,times,geometry,float,array,mathrsfs,graphics,epstopdf,latexsym,comment,cancel}

\usepackage[normalem]{ulem}
\usepackage[mathscr]{euscript}

\newcommand{\bom}[0]{{\bar{\omega}}}
\newcommand{\bga}[0]{{\bar{\gamma}}}
\newcommand{\bomga}[0]{{\bar{\omega}\bar{\gamma}}}

\begin{document}

%\title{Path superposition as resource for perfect quantum teleportation with separable states}

\title{Path superposition activating perfect quantum teleportation ability for separable states}
% ~\textcolor{red}{[should we change the title? specially "perfect quantum teleportation"...should write maybe "perfect probabilistic quantum teleportation" because of referee 2 Q. no 2]}}

\author{Sayan Mondal}
% \email{sayanmondal@hri.res.in}

\author{Priya Ghosh}
% \email{priyaghosh@hri.res.in}

\author{Ujjwal Sen}
% \email{ujjwal@hri.res.in}

\affiliation{Harish-Chandra Research Institute, Chhatnag Road, Jhunsi, Prayagraj  211 019, India\\
Homi Bhabha National Institute, Training School Complex, Anushakti Nagar, Mumbai 400 094, India}

\begin{abstract}
Quantum teleportation is a quintessential quantum communication protocol that enables the transmission of an arbitrary quantum state between two distant parties without physically transmitting the state with the help of shared entanglement and limited classical communication. We show that it is possible to relax the entanglement requirement in quantum teleportation if we have access to a certain strain of superposition of quantum processes.  Two types of superposition of quantum processes are generally considered in the literature: superposition of paths identified with quantum maps and superposition of indefinite causal orders of the maps. We find that when superposition of paths is incorporated in the protocol, quantum teleportation with unit fidelity becomes possible —  with nonzero probability of $1/4$ — even when the two parties share certain classes of separable states, including pure product states. In contrast, the assistance of superposition of indefinite causal order of quantum maps in teleportation protocol does not enable any quantum advantage for shared pure product states. Furthermore, we show that separable Werner states can also yield quantum advantage in quantum teleportation assisted by the superposition of paths. Finally, we establish that the presence of quantum coherence in the control qubit is both necessary and sufficient to achieve quantum advantage in quantum teleportation assisted with superposition of paths. The results potentially uncover yet another role of  quantum superposition, in general, in  teleportation versus entanglement. 
\end{abstract} 

\maketitle

\section{Introduction}
Quantum teleportation~\cite{Bennett1993, Pirandola2015} is one of the most striking quantum phenomena in quantum information, where a state is teleported from one party, say Alice, to another distant party, say Bob,  without physically sending any physical quantum systems. Instead, the protocol uses entangled states shared between the two parties and limited classical communication to teleport the quantum state.

%In the present day, q
Quantum teleportation is not only of fundamental significance, but also useful in applications, and for example 
%it is used to send  quantum information over long distances, but also 
plays an active role in the development of various quantum technologies like quantum repeaters~\cite{Briegel1998}, gate teleportation~\cite{Brassard1998, Gottesman1999, Knill2001, Aliferis2004, Gao2010}, 
%measurement-based computing~\cite{Raussendorf2001}, 
and many others~\cite{PhysRevLett.71.4287, Zukowski1995, Bose1998, Bose1999, Murao1999, Bennett2001}. Quantum teleportation has been experimentally realized in photonic systems~\cite{Bouwmeester1997,Boschi1998, Ursin2004}, ion traps~\cite{Barrett2004,Riebe2004}, nuclear magnetic resonances (NMR)~\cite{Nielsen1998}, and solid-state systems~\cite{Gao2013,Steffen2013, Bussieres2014, Pfaff2014}. Quantum teleportation has also been investigated for higher-dimensional systems, both in theory~\cite{Werner2001, Al-Amri2010, Wang2017, Fonseca2019, Luo2019, deSilva2021, sebastian2023} and experiments~\cite{Zhang2006, Goyal2014}. Although the quantum teleportation protocol was first introduced for discrete systems,  it has also been implemented in continuous-variable systems~\cite{Vaidman1994, Braunstein1998, Furusawa1998}.

In the standard qubit teleportation protocol, if neither classical communication nor shared randomness is allowed, the receiver cannot do better than randomly guessing the quantum state to be teleported. In this case, the average fidelity between the guessed state and the actual input state, averaged over all pure input states is ${1}/{2}$. When the sender and receiver are allowed to use classical communication and share separable states, the maximum achievable average fidelity over all protocols improves to ${2}/{3}$~\cite{Massar1995, horodecki1999}. Therefore, any qubit teleportation protocol that achieves an average fidelity greater than ${2}/{3}$ is said to demonstrate quantum advantage.
It is well established that the presence of shared entanglement together with classical communication is necessary to surpass this classical threshold. In particular, using a maximally entangled state as the shared resource allows for perfect quantum teleportation, achieving unit fidelity~\cite{Popescu1994, Massar1995, chau1996, horodecki1999, Lo1999, Henderson2000, ghosh2000}.

\textcolor{black}{Since shared entanglement is necessary to achieve a quantum advantage over the classical fidelity threshold in standard quantum teleportation, it is natural to ask whether such an advantage can be obtained using alternative quantum resources, while keeping other operations, such as local operations and classical communication, unchanged. Interestingly, recent studies have shown that the superposition of quantum processes~\cite{Oi2003, Chiribella2013, Ebler2018, Chiribella2019, abbott2020, Chiribella2021} can provide advantages in a variety of quantum information tasks, including quantum communication~\cite{Chiribella2012,hullamballi2025,kechrimparis2025, wu2025}, quantum thermodynamics~\cite{Guha2020,Felce2020,nie2022quantumcool, cao2022, nie2022, dieguez2023}, quantum games~\cite{oreshkov2012}, quantum metrology~\cite{Zhao2020, agrawal2025, Guo2025}, and others. Motivated by these observations, in our work we address the question of whether a quantum advantage in teleportation can be achieved in the complete absence of shared state entanglement, by exploiting coherent superpositions of local operation and classical communication (LOCC) teleportation maps.
}

%{\color{black} Despite this well-understood role of shared entanglement, it is natural to ask whether the classical fidelity bound can be surpassed without entanglement in the shared state, but instead by exploiting genuinely quantum features at the level of operations. Recent developments in quantum information theory suggest that quantum advantage need not always originate from state entanglement alone, but may also arise from nonclassical control over the dynamical structure of quantum processes themselves. In particular, allowing different quantum operations to be applied in coherent superposition introduces a new operational resource that has no classical analogue. This raises a fundamental question: can superposition of quantum processes compensate for, or even partially replace, the role traditionally played by shared entanglement in quantum teleportation? 
% Addressing this question is essential both for understanding the minimal resources required for teleportation beyond the classical limit and for identifying new forms of quantum advantage rooted in process-level coherence rather than state entanglement.}

%In more detail, the superposition of quantum processes has recently gained significant attention in quantum information science~\cite{Oi2003, Chiribella2013, Ebler2018, Chiribella2019, abbott2020, Chiribella2021}. In this framework, 
\textcolor{black}{In order to realize superposition of quantum processes,
an additional quantum system, known as the control system, is introduced. }
%to realize superposition of quantum processes.
For example, if the control system is in a state, say $\ket{0}$, a process $\mathcal{P}_1$ is applied; if it is in a state, say $\ket{1}$, another process $\mathcal{P}_2$ is applied. When the control system is prepared in a quantum superposition of these states $\ket{0}$ and $\ket{1}$, the evolution results in a coherent superposition of the corresponding quantum processes. Thus, one can realize the superposition of any two quantum processes. This idea can naturally be extended to more than two quantum processes, enabling superposition over any finite number of quantum operations using a higher-dimensional control system.

The superposition of quantum processes can generally be categorized into two cases. One is the superposition of paths: to coherently superpose two or more quantum channels~\cite{Oi2003, Chiribella2019, abbott2020}. And another is the superposition of indefinite causal order: to superpose the causal order of two or more quantum channels~\cite{Chiribella2013, Ebler2018, Chiribella2021}, commonly referred to as a quantum switch. 
The coherent superposition of quantum processes has been investigated in photonic systems in experiments~\cite{Procopio2015, Rubino2017, Goswami2018, Rubino2021, Rubino2022, Stromberg2024}; for a review, see~\cite{Rozema2024}. More recently, this has been theorized for generic quantum systems~\cite{Ban2020, BAN2021, Siltanen2021, Lin2022}.
% and has been shown to be advantageous in various quantum applications.
% , including quantum communication~\cite{ hullamballi2025}, quantum thermodynamics~\cite{Felce2020,nie2022quantumcool}, and so on.
% Moreover, the quantum switch has been shown to be useful in various tasks such as quantum communication~\cite{Chiribella2012, kechrimparis2025, wu2025}, quantum games~\cite{oreshkov2012}, thermodynamic advantages~\cite{Guha2020, cao2022, nie2022, dieguez2023}, and quantum metrology~\cite{Zhao2020, agrawal2025, Guo2025}. 
Additionally, it is also possible to enhance the non-Markovianity of quantum channels using quantum switches~\cite{ghosh2024}.

%Entanglement is a necessary resource to teleport any arbitrary quantum state between two spatially separated parties in the standard teleportation protocol. In this work, we investigate whether the requirement of entanglement persists when the superposition of quantum processes is included in the standard quantum teleportation protocol. Specifically, we consider two widely studied forms of superposition of quantum processes: superposition of paths and superposition of indefinite causal orders of quantum maps.
% \textcolor{black}
{As a result, our findings show that} when a superposition of paths is incorporated into the standard quantum teleportation protocol, quantum advantage can be achieved even when the shared states are separable. In particular, we find that for a certain class of separable states, including pure product states, teleportation with unit fidelity can be realized with a success probability of ${1}/{4}$. In contrast, we find that employing a superposition of indefinite causal orders in the teleportation protocol does not provide any quantum advantage in quantum teleportation when the shared states are pure product states.
We further analyze two-qubit Werner states,
which are mixtures of the maximally entangled state (with mixing probability $p$) and the maximally mixed state. Such states 
are separable for mixing probabilities $p \leq {1}/{3}$. We find that quantum teleportation assisted by a superposition of paths exhibits quantum advantage for $p > {1}/{5}$.
Lastly, we establish that coherence in the control qubit, which governs the superposition of quantum processes, is both necessary and sufficient to achieve the quantum advantage in quantum teleportation for shared separable states. Thus, while entanglement is indispensable in the standard teleportation setting, superposition of paths acts as an alternative quantum resource enabling  {\color{black}  unit-fidelity teleportation, with nonzero probability conditional on measurement outcomes, even when separable states are being shared.}

% {\color{black} In quantum mechanics, superposition of local orthogonal states generates entanglement and thereby quantum correlations. 
% Analogously, it is now well established that superposing different quantum channels can provide operational advantages in quantum information tasks that are unattainable without such superposition~\cite{Chiribella2021,abbott2020,Rubino2021}. 
% These advantages originate from correlations created by the coherent superposition of channels, which we refer to as ``process entanglement". In our scheme, we exploit a superposition of LOCC channels to implement quantum teleportation and demonstrate that, even in the complete absence of shared state entanglement, unit-fidelity teleportation can be achieved probabilistically due to the emergence of process entanglement. We show that nonzero quantum coherence in the control system is both necessary and sufficient to obtain a quantum advantage within our protocol, thereby establishing process entanglement as an operational resource for teleportation. Unlike conventional entanglement-assisted teleportation~\cite{Massar1995,horodecki1999}, the advantage here does not rely on shared entangled states, offering a complementary and conceptually distinct perspective on the resource requirements of quantum teleportation.
% }
{\color{black}In quantum mechanics, the coherent superposition of locally orthogonal states can generate entanglement and, consequently, quantum correlations. 
By analogy with state entanglement, it has now been firmly established that the coherent superposition of distinct quantum channels can provide operational advantages in quantum information processing tasks that are unattainable in the absence of such superposition~\cite{Chiribella2021,abbott2020,Rubino2021}. 
% {\color{blue}These advantages stem from coherent superposition of channels induced by the quantum coherence in the control qubit.}
% correlations induced by the coherent control of channels.}
% , which we term ``process entanglement.''
% {\color{blue}In this work, we exploit a coherent superposition of LOCC channels to implement quantum teleportation and demonstrate that, even in the complete absence of shared state entanglement, unit-fidelity teleportation can be achieved probabilistically.}
% due to the emergence of process entanglement. 
% {\color{blue}We further prove that nonzero quantum coherence in the control system is both necessary and sufficient to obtain a quantum advantage within our protocol.}
% , thereby establishing process entanglement as a genuine operational resource for teleportation. 
These advantages stem from correlations induced by the coherent control of channels, which we term ``process entanglement.”In this work, we exploit a coherent superposition of LOCC channels to implement quantum teleportation and demonstrate that, even in the complete absence of shared state entanglement, unit-fidelity teleportation can be achieved probabilistically due to the emergence of process entanglement. We further prove that nonzero quantum coherence in the control system is both necessary and sufficient to obtain a quantum advantage within our protocol, supporting the interpretation of process entanglement as an operational resource.
Unlike conventional entanglement-assisted teleportation~\cite{Massar1995,horodecki1999}, where the advantage derives from shared entangled states, the enhancement demonstrated here arises solely from the superposition of channels. 
Our results therefore provide a complementary and conceptually distinct perspective on the resource requirements of quantum teleportation.}

% {\color{black}
% While {conventional teleportation} requires entanglement between the sender and receiver to achieve a quantum advantage, 
% interestingly, we show that in the complete absence of state entanglement, the superposition of paths, through the emergence of process entanglement, can enable unit-fidelity probabilistic quantum teleportation, and the nonzero amount of  quantum coherence in the control system is both necessary and sufficient to achieve a quantum advantage in teleportation within our protocol. More specifically, constructing a superposition of channels requires a nonzero amount of quantum coherence in the control system, which gives rise to process entanglement. 
% Taken together, these facts establish that process entanglement acts as an operational resource in our quantum teleportation protocol.
% Our results thus demonstrate that the quantum resource responsible for teleportation fidelity can be shifted from the shared state to the dynamical structure of the protocol itself, offering a new perspective on the role of quantum resources in teleportation.
% }
{
% {\color{blue}While conventional teleportation requires shared entanglement between the sender and receiver to achieve a quantum advantage, we demonstrate that, even in the complete absence of state entanglement, a coherent superposition of paths can enable unit-fidelity probabilistic teleportation.}
% through the emergence of process entanglement. 
% {\color{blue}Specifically, implementing a superposition of channels necessitates nonzero quantum coherence in the control system.}
While conventional teleportation requires shared entanglement between the sender and receiver to achieve a quantum advantage, we demonstrate that, even in the complete absence of state entanglement, a coherent superposition of paths can enable unit-fidelity probabilistic teleportation through the emergence of process entanglement. Specifically, implementing a superposition of channels necessitates nonzero quantum coherence in the control system, which in turn generates process entanglement. 
% , which in turn generates process entanglement. 
We prove that this coherence is both necessary and sufficient for achieving a quantum advantage within our protocol. 
% These results establish process entanglement as an operational resource for quantum teleportation. 
More broadly, our results show that the resource underlying teleportation fidelity need not reside in a shared entangled state, but can instead be encoded in the coherence of the control qubit, which enables a coherent superposition of quantum channels. This provides a conceptually distinct perspective on the role of quantum resources in teleportation.
}
{\color{black}{We emphasize that the present protocol does not constitute a fundamentally new teleportation mechanism. Rather, it may also be viewed as a conditional implementation of a quantum channel, selected via measurement of the control qubit. The central feature of the protocol is the coherent superposition of distinct LOCC maps enabled by quantum coherence in the control system. This coherent control gives rise to interference terms that are absent in any classical mixture of operations, even in the presence of post-selection, and it is this genuinely quantum interference effect that underlies the observed advantage.
% We note that the protocol can be viewed operationally as a conditional implementation of a quantum channel, selected via measurement on the control qubit. However, its essential feature lies in the coherent superposition of maps, enabled by quantum coherence in the control system. This generates interference terms, which are absent in any classical mixture of LOCC operations, even with post-selection. In the absence of such coherence, the protocol reduces to a classical mixture and yields no quantum advantage.
}}

The remaining part of the paper is organized as follows. In Sec.~\ref{set-up}, we provide a brief overview of the standard quantum teleportation protocol and introduce the framework of superposition of quantum processes. In Sec.~\ref{sec3}, we study standard teleportation protocol enhanced by superposition of quantum processes. This section presents our main results, demonstrating that a superposition of paths can enable probabilistic quantum teleportation with unit-fidelity, even when the shared states are separable, including pure product states. Furthermore, in this section, we show that employing a superposition of indefinite causal orders does not offer any quantum advantage in the standard teleportation protocol when the shared states are pure product and a class of separable two-qubit Werner states also provides quantum advantage in quantum teleportation when assisted by a superposition of paths. 
In Sec.~\ref{sec:role-coherence}, we show that the coherence of the initial state of the control qubit, is both necessary and sufficient for activating shared separable states for quantum teleportation. Finally, we conclude in Sec.~\ref{conclu}.

\begin{figure*}
    \centering
    \includegraphics[width=0.75\linewidth]{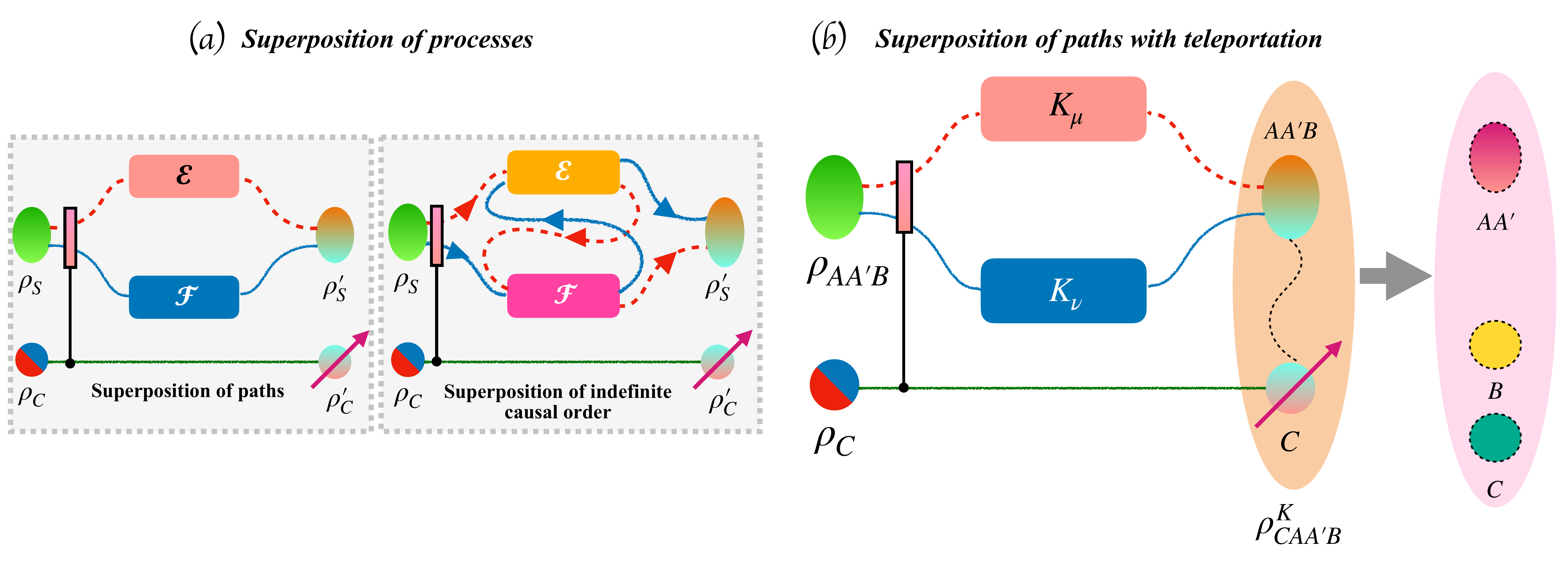}
\caption{\color{black}{\textbf{\emph{Schematic diagram of  superposition of quantum processes and  implementation of teleportation protocol via the
superposition of paths}}.
(a) The superposition of quantum processes on a target system is realized via a control system $C$. The target can undergo one of two quantum processes: $\mathcal{P}_1$  or $\mathcal{P}_2$ , determined by the control qubit. If the control is in a state $\ket{0}$ (represented by blue), a process $\mathcal{P}_1$ is applied; if it is in the state $\ket{1}$ (represented by blue), $\mathcal{P}_2$ is applied. In the figure the two processes $\mathcal{P}_1$ and $\mathcal{P}_2$ are represented by red dotted line and blue solid line respectively. The rectangular pink box represents the conditional implementation of either $\mathcal{P}_1$ or $\mathcal{P}_2$ depending on the state of $C$.  By preparing the control in a coherent superposition $\ket{+}_C \coloneqq {(\ket{0} + \ket{1})}/{\sqrt{2}}$, a superposition of these two processes can be acted on the target. The state $|+\rangle_C$ is represented by an equally split red–blue disk, indicating an equal superposition of the two basis states.
The superposition of processes consists broadly of two kinds of situations - superposition of paths and superposition of indefinite causal order.  
In superposition of paths, the two processes are two maps, $\mathcal{P}_1 = \mathcal{E}$ and $\mathcal{P}_2 = \mathcal{F}$, represented by the red and blue boxes respectively. In superposition of indefinite causal order of maps, the two processes are the order of the maps - $\mathcal{P}_1 = \mathcal{E}\circ\mathcal{F}$ and $\mathcal{P}_2 = \mathcal{F}\circ\mathcal{E}$. The control qubit is measured in the end, which collapses the state of the system to a specific state depending on the outcome of the measurement. (b) We present the schematic for the implementation of the teleportation protocol with the superposition of paths. There are three qubits, $A$ which is in the state that is to be teleported, $B$ where the state is to be teleported and $A^\prime$. In conventional teleportation scheme, $A$ and $A^\prime$ are kept together and are spatially separated from $B$. For teleportation, in the conventional teleportation protocol, $A^\prime$ and $B$ need to share an entangled state. Only local operations and classical communications are allowed in the three parties. 
In the scheme introduced in the current work, an extra control qubit $C$ is also introduced. The two processes here are the maps corresponding to the teleportation protocol as presented in Eq.~(2). We obtain a state $\rho_{CAA'B}^K$ given in Eq.~(7). The control qubit is measured and the state that is to be teleported from $A$ is probabilistically obtained at $B$, even if $A^\prime$ and $B$ are separable as is necessary for conventional teleportation protocol.   
}}
\label{fig:schematic}
\end{figure*}

\section{The Setup}
\label{set-up}
% Quantum teleportation requires shared entanglement and classical communication between the two spatially separated parties. 
% In this work, we investigate whether entanglement between the two distant parties is  necessary to obtain quantum advantage in quantum teleportation when superposition of quantum processes is applied in conjunction with the standard teleportation protocol. We keep our discussions limited to the teleportation of pure qubit states throughout this paper.
% Quantum teleportation conventionally requires shared entanglement and classical communication between spatially separated parties. 
We examine whether entanglement is necessary for achieving quantum advantage when the standard teleportation protocol is supplemented with superposition of quantum processes. 
In this section we discuss the usual quantum teleportation protocol and the two types of superposition of processes. Throughout this paper, $\mathbb{I}_d$ denotes the identity operator acting on the Hilbert space of dimension $d$. Throughout, we restrict our analysis to the teleportation of pure qubit states.
Here $\sigma_i \coloneqq \{\sigma_x,\sigma_y, \sigma_z\}$ denote Pauli operators and $|\phi^\pm\rangle,|\psi^\pm\rangle$ represent the Bell states.
$\psi^\pm$ and $\phi^\pm$ denote $\ket{\psi^\pm}\bra{\psi^\pm}$ and $\ket{\phi^\pm}\bra{\phi^\pm}$ respectively.

\subsection{Quantum teleportation protocol}
Let us suppose that Alice has a two subsystems denoted by $A$ and $A^\prime$, whose states belong to the Hilbert spaces  $\mathcal{H}_A$ and $\mathcal{H}_{A^\prime}$ respectively. On the other hand, Bob has one subsystem denoted by $B$ whose states belong to Hilbert space $\mathcal{H}_B$. Alice wants to teleport any pure state in $\mathcal{H}_A$ to Bob, using a shared state in $\mathcal{H}_{A^\prime} \otimes \mathcal{H}_B$. In this work, we consider all of $\mathcal{H}_A$, $\mathcal{H}_{A^\prime}$ and $\mathcal{H}_B$ to be of dimension two.  
In the standard quantum teleportation protocol~\cite{Bennett1993}, Alice wants to send a quantum state, let us say $|\phi\rangle_A \langle \phi|$ to Bob, who is at a distant location from her, using a shared maximally entangled state $|\beta\rangle_{A^\prime B }$  on $\mathcal{H}_{A^\prime} \otimes \mathcal{H}_{B}$ and local operations and one-way classical communication by Alice and Bob.  Here $|\beta\rangle$ can be any one of the Bell states. Hence, the initial joint state with Alice and Bob is of the form $\rho_{AA^\prime B} ( \coloneqq |\phi\rangle_A \langle \phi|\otimes|\beta\rangle_{A^\prime B}\langle\beta|) $. 
Let us briefly discuss the \textit{standard teleportation protocol} as follows.

\begin{itemize}
    \item Alice does a Bell basis measurement onto $\{|\phi^\pm\rangle,|\psi^\pm\rangle\}$ on her parties $AA^\prime$.

    \item Alice sends her outcome to Bob using classical communication. Bob applies local unitary operations on his party $B$ depending on the information Alice sent him. 

    \item To make calculations easier, we add another step in the protocol. Alice applies a unitary operator on one of her qubits post-measurement, based on her measurement outcome, to prepare it in the initially shared maximally entangled state. This does not affect the teleportation protocol since it is just a local operation at one of Alice's subsystem. 
\end{itemize}

It has been shown that the standard teleportation protocol teleports the exact state from Alice to Bob~\cite{Bennett1993}. The overall transformation induced by the protocol on the joint Alice-Bob state acting on Hilbert space $\mathcal{H}_A \otimes \mathcal{H}_{A^\prime} \otimes \mathcal{H}_B$ can be described in terms of Kraus operator representation as follows:
\begin{align}
    \rho_{AA^\prime B}^\prime = (|\beta\rangle \langle \beta|)_{AA^\prime} \otimes (|\phi\rangle \langle \phi|)_B =\sum_{i = 0}^3 T_i\rho_{AA^\prime B} T_i^\dagger,
\end{align}
where $\sum_i T_i^\dagger T_i = \mathbb{I}_8$.
When  $|\beta\rangle \coloneqq |\psi^-\rangle$ is considered as shared state in this standard teleportation protocol, the Kraus operators are given by 
\begin{align}
    K_0 &\coloneq (|\psi^-\rangle \langle \psi^-|)_{AA^\prime} \otimes (\mathbb{I}_2)_B, \nonumber \\
    K_1 &\coloneq (|\psi^-\rangle \langle \psi^+|)_{AA^\prime} \otimes (\sigma_z)_B, \nonumber\\
    K_2 &\coloneq (|\psi^-\rangle \langle \phi^-|)_{AA^\prime} \otimes (\sigma_x)_B, \nonumber \\
    K_3 &\coloneq (|\psi^-\rangle \langle \phi^+|)_{AA^\prime} \otimes (i\sigma_y)_B, 
    \label{psi-kraus}
\end{align}
with $\sum_i K_i^\dagger K_i = \mathbb{I}_8$ .
When the shared state is $|\beta\rangle \coloneqq |\phi^-\rangle$, the Kraus operators will have the following form:
\begin{align}
    L_0 &\coloneq (|\phi^-\rangle \langle \phi^-|)_{AA^\prime} \otimes (\mathbb{I}_2)_B, \nonumber \\
    L_1 &\coloneq (|\phi^-\rangle \langle \phi^+|)_{AA^\prime} \otimes (\sigma_z)_B, \nonumber\\
    L_2 &\coloneq (|\phi^-\rangle \langle \psi^-|)_{AA^\prime} \otimes (\sigma_x)_B,\nonumber \\
    L_3 &\coloneq (|\phi^-\rangle \langle \psi^+|)_{AA^\prime} \otimes (i\sigma_y)_B, 
    \label{phi-kraus}
\end{align}
with $\sum_i L_i^\dagger L_i = \mathbb{I}_8$.
\par
In the remaining part of the paper, we will only consider the above standard teleportation protocol in conjunction with either superposition of paths or with superposition of indefinite causal orders of maps.

\subsection{Superposition of quantum processes}
The superposition of quantum processes generally falls into two categories: superposition of paths and superposition of indefinite causal order of quantum maps. In both cases, a quantum state can evolve either through \emph{Process~1}~($\mathcal{P}_1$) or \emph{Process~2}~($\mathcal{P}_2$). Throughout our paper, we consider both $\mathcal{P}_1$ and $\mathcal{P}_2$ to be completely positive and trace-preserving (CPTP) maps.
To realize the superposition of these quantum processes, an additional control system, denoted by $C$ is introduced. Depending on the state of the control system, either $\mathcal{P}_1$ or $\mathcal{P}_2$ is applied to the target system. When the control system is prepared in a coherent superposition of the basis states, it enables a coherent superposition of the corresponding quantum processes.
A schematic illustration of the superposition of quantum processes is shown in Fig.~\ref{fig:schematic}. 
% where $\rho_{AA^\prime B}$ and $\ket{\chi}_C\bra{\chi}$ represent the state of the target system and the control system, respectively. 
Here, the dimension of Hilbert space of the control system is chosen to be two. Let us denote the Hilbert spaces of target and control systems by $\mathcal{H}_S$ with dimension $d_S$ and $\mathcal{H}_C$ with dimension $d_C$, respectively.

\emph{\textbf{Superposition of paths}}~\cite{Oi2003, Chiribella2019, abbott2020}. 
In superposition of quantum maps, the processes $\mathcal{P}_1$ and $\mathcal{P}_2$ are individual CPTP maps that act on the system depending on the state of the control system. 
Since quantum mechanics allows superposition of states, it is possible to create a superposition of the two maps using a realization of the superposition of quantum states of an additional control system. 
The Kraus operator representation of superposition of quantum maps $\mathcal{P}_1$ and $\mathcal{P}_2$ can be written as
\begin{align*}
    \mathcal{M}_{\mu \nu} \coloneqq \ket{0}_C\langle 0|\otimes \mathcal{E}_\mu + \ket{1}_C\langle 1|\otimes \mathcal{F}_\nu,
\end{align*}
where $\{|0\rangle, |1\rangle\}$ represents a basis of the Hilbert space of the control system. Here
$\{\mathcal{E}_\mu\}$ and $\{\mathcal{F}_\nu\}$ represent Kraus operators of the maps in $\mathcal{P}_1$ and $\mathcal{P}_2$ respectively with $\sum_\mu \mathcal{E}_\mu^\dagger \mathcal{E}_\mu = \mathbb{I}_{d_S}$ and $\sum_\nu \mathcal{F}_\nu^\dagger \mathcal{F}_\nu = \mathbb{I}_{d_S}$. 

\emph{\textbf{Superposition of indefinite causal order of maps}}~\cite{Chiribella2013,  Ebler2018, Chiribella2021}. In superposition of indefinite causal order of two quantum maps, the processes $\mathcal{P}_1$ and $\mathcal{P}_2$ are composed of two CPTP maps, but with their causal order being reversed in $\mathcal{P}_1$, with respect to $\mathcal{P}_2$. 
The processes can be written as $\mathcal{P}_1 \coloneqq \mathcal{E}\circ \mathcal{F}$ and $\mathcal{P}_2 \coloneqq \mathcal{F} \circ \mathcal{E}$.  The Kraus operator representation of the superposition of indefinite causal order of two quantum maps, $\mathcal{P}_1$ and $\mathcal{P}_2$ can be written as
\begin{align*}
    \mathcal{M}_{\mu \nu} \coloneqq \ket{0}_C\langle 0|\otimes \mathcal{E}_\mu \mathcal{F}_\nu + \ket{1}_C\langle 1|\otimes \mathcal{F}_\nu\mathcal{E}_\mu,
\end{align*}
where $\{\mathcal{F}_\nu\}$ and $\{\mathcal{E}_\mu\}$ are the Kraus operators of the maps $\mathcal{F}$ and $\mathcal{E}$ respectively. 

Let us denote initial control and target systems by $|\chi\rangle_C\langle\chi|$ and $\rho_S$  respectively.
Thus, the joint state of the target-control system, $\rho_\text{CS} (\coloneqq |\chi\rangle_C\langle\chi|\otimes \rho_S)$, will be transformed under the action of the superposition of quantum processes as follows:
\begin{align*}
    \rho^\prime_{CS} \coloneqq \sum_{\mu,\nu} \mathcal{M}_{\mu\nu} |\chi\rangle_C\langle\chi|\otimes \rho_S \mathcal{M}^\dagger_{\mu\nu},
\end{align*}
where $\sum_{\mu,\nu} \mathcal{M}_{\mu\nu} \mathcal{M}^\dagger_{\mu\nu} = \mathbb{I}_{d_C d_S}$. Here $\mathbb{I}_{d_C d_S}$ denotes the identity operator on Hilbert space $\mathcal{H}_{C} \otimes \mathcal{H}_{S}$.

We are now ready to present the results of our paper, which are detailed in the following sections.

\section{Teleportation powered by superposition of quantum processes}
\label{sec3}
% In the standard quantum teleportation protocol, success is usually quantified using the average fidelity of the teleported state with the initial state that was supposed to be teleported. Since, in this paper we only consider teleportation of pure single-qubit states, we shall consider averaging the fidelity between the input and output state over all input pure states.
% In the standard teleportation protocol, the average fidelity turns out to be $\frac{1}{2}$ if the output state is randomly guessed.
% Classically, the optimal protocol can achieve an average fidelity of $\frac{2}{3}$. Additionally, the standard teleportation protocol also achieves a maximum of $\frac{2}{3}$ average fidelity when separable states are shared between Alice and Bob. Thus, quantum advantage of teleportation is quantified by the average fidelity, $>\frac{2}{3}$ and it is obtained when the shared state is entangled. As a special case, any state can be perfectly teleported with unit fidelity when the shared state is maximally entangled. 
% Thus, entanglement is necessary for obtaining quantum advantage in standard teleportation protocol.
% Teleportation performance is quantified by the average fidelity, taken here over all pure single-qubit input states. In the standard protocol, both the optimal classical strategy and any protocol using separable shared states are bounded by an average fidelity of $2/3$, while fidelities exceeding this value require entanglement, with unit fidelity achieved for maximally entangled resources.
In this section we investigate if entanglement of the shared state is necessary for quantum advantage when we apply either the superposition of paths or superposition of indefinite causal order of maps in the standard teleportation protocol.
For this, we mainly consider the two-qubit diagonal (with respect to the local computational basis) separable states  and two-qubit separable Werner states presented in Eqs.~\eqref{state1} and~\eqref{werner-state} respectively as the shared states in our teleportation protocol. 

{\color{black}In the standard quantum teleportation protocol, entanglement of the shared bipartite state between the sender and the receiver is an indispensable resource; in the absence of shared entanglement, the achievable teleportation fidelity is bounded by the classical limit. In contrast, the protocol considered here relies on a fundamentally different mechanism. The quantum advantage does not stem from entanglement in the shared state, which may even be fully separable, but rather from the coherent superposition of distinct quantum processes controlled by an auxiliary qubit. The coherence of the control system enables interference between the paths, generating nonclassical correlations at the level of the implemented quantum map. 
Consequently, the relevant resource is shifted from static state entanglement to dynamical process entanglement arising from coherent control.
% {\color{blue}Consequently, the relevant resource is shifted from entanglement in the shared state to quantum coherence of the control qubit.}
This shift allows  teleportation {of unit-fidelity} with nonzero probability even when the shared state is separable, thereby explicitly relaxing the entanglement requirements on the shared resource compared to the standard teleportation scheme.}

\vspace{-5 mm}
\subsection{Teleportation with assistance of superposition of paths when shared states are diagonal in local basis}
\label{diagonal-sec}
A class of two-qubit states that are diagonal in the local computational basis of $\{|00\rangle, |01\rangle, |10\rangle, |11\rangle\}$, can be written as
\begin{align}
\rho_{A^\prime B} \coloneqq
\begin{pmatrix}
p_0 & 0 & 0 & 0 \\
0 & p_1 &0 & 0 \\
0 & 0 & p_2 & 0 \\
0 & 0 & 0 & p_3
\end{pmatrix},
\label{state1}
\end{align}
with $p_i \geq 0 \hspace{1 mm} \forall i \in \{0,1,2,3\}$ and $\sum_{i=0}^3 p_i = 1$. These states are separable states. Here, we consider the shared state to be of the form of Eq.~\eqref{state1} and apply the standard teleportation protocol in conjunction with superposition of paths to investigate if such a separable shared states can provide quantum advantage.

Let us consider any arbitrary pure single-qubit state $|\phi\rangle$ that we want to teleport from Alice to Bob using the shared state $\rho_{A^\prime B}$ written in Eq.~\eqref{state1}. We prepare the control qubit $C$ in the state $|\chi\rangle_C \coloneqq \alpha \ket{0} + \beta \ket{1}$ with $|\alpha|^2 + |\beta|^2 = 1$. Hence, the initial joint control-Alice-Bob state is given by
\begin{align}
    \rho_{CAA^\prime B} \coloneqq |\chi\rangle_C\langle\chi|\otimes  |\phi\rangle_A\langle\phi| \otimes \rho_{A^\prime B}.
    \label{eq:rhoCAAB}
\end{align}

{\color{black}Quantum teleportation via a superposition of paths has previously been investigated in Ref.~\cite{Sen_2025}, where the two paths correspond to the presence or absence of a maximally entangled shared resource. In that work, one path implements the conventional teleportation protocol using a maximally entangled state, while the alternative path corresponds either to the identity operation or to a purely classical communication channel. In both scenarios, the protocol relies on the availability of a maximally entangled state.
In contrast, the protocol introduced here enables probabilistic teleportation using shared separable states, including product states, and therefore does not require shared entangled state as a resource. Furthermore, our scheme achieves unit-fidelity teleportation with nonzero probability, a feature that is not present in the protocol of Ref.~\cite{Sen_2025}, which does not attain unit fidelity even probabilistically.}

% In Ref.~\cite{Sen_2025}, the standard teleportation of quantum state using superposition of paths was considered, where the two paths were availability and unavailability of maximally entangled state.
% {\color{teal} 
% Two distinct scenarios were analyzed. In the first, one path implements the standard teleportation protocol using a maximally entangled resource, while the other path corresponds to the identity operation, effectively leaving the input state unchanged. In the second scenario, one path again realizes the standard teleportation protocol with a maximally entangled state, whereas the alternative path consists solely of classical communication. Unlike our protocol, which achieves unit fidelity with a nonzero probability, their scheme does not attain unit-fidelity teleportation, even probabilistically. ~\textcolor{red}{[not clear yet!!]}
% In this work, the authors considered two cases. In the first case, they considered the first path to be the conventional teleportation protocol (with a maximally entangled shared state) and the other path is identity operator that is doing nothing. The second case, the first path considered is the conventional teleportation protocol (with a maximally entangled shared state) and the other path consists of classically communication.
% }

\textcolor{black}{Here we apply conventional teleportation protocol in conjunction with superposition of paths as follows.
We construct a superposition of LOCC teleportation maps that arise in the conventional teleportation protocol when the shared state is maximally entangled.
In particular, the joint system $AA'B$ undergoes a coherent superposition of two teleportation maps represented by Kraus operators $\{K_\mu\}$ and $\{K_\nu\}$, each corresponding to the conventional teleportation protocol with the shared state $|\beta\rangle \equiv |\psi^-\rangle$, as given in Eq.~\eqref{psi-kraus}. This superposition is implemented by a control system: the system $AA'B$ 
undergoes a path-dependent evolution, governed by $\{K_\mu\}$ when the control qubit $C$ is in $\ket{0}$ and by $\{K_\nu\}$ when $C$ is in $\ket{1}$, thereby realizing a coherent superposition of quantum processes along two interferometric paths. 
Eq.~\eqref{map-K} provides the explicit form of the Kraus operators $M_{\mu\nu}^K$ implementing this coherent superposition, while Eq.~\eqref{eq:total_operation} gives the resulting output state $\rho^{K}_{CAA'B}$ obtained from an arbitrary input state $\rho_{CAA'B}$.
% {\color{black} The teleportation protocol is given by the Kraus operators given in Eq.~\eqref{psi-kraus}. The joint state of $AA'B$ undergoes coherent superposition of the Kraus operator $\{K_\mu\}$ and $\{K_\nu\}$ in the two individual paths. The system of $AA'B$ undergoes $\{K_\mu\}$ when the control qubit $C$ in in the state $|0\rangle$ while, it undergoes $\{K_\nu\}$ when $C$ is in the state $|1\rangle$. }
%{\color{black} The teleportation protocol is described by the Kraus operators in Eq.~\eqref{psi-kraus}. The joint system $AA'B$ undergoes a path-dependent evolution, governed by $\{K_\mu\}$ when the control qubit $C$ is in $\ket{0}$ and by $\{K_\nu\}$ when $C$ is in $\ket{1}$, as implemented by the operator $M_{\mu\nu}^K$. When the control qubit is prepared in a coherent superposition of $\ket{0}$ and $\ket{1}$, this conditional dynamics generates a coherent superposition of the two teleportation processes in the interferometer arms.}
%Thus, the Kraus operators $\{M^K_{\mu\nu}\}$ is defined as
The explicit form of the Kraus operators $M_{\mu\nu}^K$ implementing this coherent superposition is given by}
\begin{align}
     M_{\mu \nu}^K \coloneqq \ket{0}\langle0|\otimes\frac{1}{2}K_\mu  + \ket{1}\langle1|\otimes\frac{1}{2}K_\nu,
     \label{map-K}
\end{align}
where the Kraus operators $\{K_i\}$ is taken from Eq.~\eqref{psi-kraus}.

%After the application of our protocol using the Kraus operators $\{M_{\mu\nu}^K\}$ defined in Eq.~\eqref{map-K} on $\rho_{CAA^\prime B}$ defined in Eq.~\eqref{eq:rhoCAAB}, we obtain
\textcolor{black}{After applying the coherent superposition of quantum processes, implemented by the operators $M_{\mu\nu}^K$, to an arbitrary input state $\rho_{CAA’B}$, the resulting state is given by}
\begin{widetext}
\begin{align}
    \rho_{CAA^\prime B}^K \coloneqq &\sum_{\mu,\nu}M_{\mu\nu}^K\rho_{CAA^\prime B}(M_{\mu\nu}^K)^\dagger \nonumber \\
    = &|\alpha|^2 \ket{0}_C\langle0|\otimes \sum_\mu K_\mu \rho_{AA^\prime B}K_\mu^\dagger +  \frac{\alpha \beta^*}{4} \ket{0}_C\langle1|\otimes \sum_{\mu,\nu} K_\mu \rho_{AA^\prime B}K_\nu^\dagger \nonumber \\ & +\frac{\alpha^* \beta}{4} \ket{1}_C\langle0|\otimes \sum_{\mu,\nu} K_\nu \rho_{AA^\prime B}K_\mu^\dagger  +|\beta|^2 \ket{1}_C\langle1|\otimes \sum_\nu K_\nu \rho_{AA^\prime B}K_\nu^\dagger,
    \label{eq:total_operation}
\end{align}
\end{widetext}
where $\rho_{AA^\prime B} \coloneqq |\phi\rangle_A\langle\phi| \otimes \rho_{A^\prime B}$.
{\color{black} In Eq.~\eqref{eq:total_operation}, we note that if the control qubit $C$ were either in the state $|0\rangle$ or $|1\rangle$, $M_{\mu\nu}^K$ would have implemented the {conventional teleportation protocol}. But as $C$ is in superposition of the two states, we obtain a coherent superposition of the two maps.}

Now, we operate the Hadamard gate on the control qubit $C$ of the final joint control-Alice-Bob state $\rho_{CAA^\prime B}^K$ and then measure the control qubit $C$ in the computational basis $\{\ket{0}, \ket{1}\}$. 
Consequently, the final state of the control qubit is in the state either $\ket{0}\!\bra{0}$ or $\ket{1}\!\bra{1}$, the corresponding final state obtained at Bob is given by $\rho_B^{\pm}$ respectively where 
\begin{align}
    {\rho}^+_B &\coloneqq \frac{2y\rho_d + 2(1-y)\sigma_x\rho_d\sigma_x  + Xy \sigma_z|\phi\rangle\langle\phi|\sigma_z}{2 + Xy} \nonumber \\
    {\rho}^-_B &\coloneqq  \frac{2y\rho_d + 2(1-y)\sigma_x\rho_d\sigma_x  - Xy \sigma_z|\phi\rangle\langle\phi|\sigma_z}{2 - Xy},
\end{align}
with $y \coloneq p_1 + p_2$, $X \coloneq \alpha\beta^* + \alpha^*\beta$, and $\rho_d \coloneq \frac{1}{2}(|\phi\rangle\langle\phi| + \sigma_z |\phi\rangle\langle\phi| \sigma_z)$. 
The corresponding probabilities to obtain the control qubit in the state $\ket{0}\!\bra{0}$ and $\ket{1}\!\bra{1}$ are as follows:
\begin{align*}
    p_+ \coloneqq \frac{2 + Xy}{4}, \hspace{4 mm}p_- &\coloneqq \frac{2 - Xy}{4},
\end{align*}
respectively.~Detailed calculations are presented in Appendix~\ref{sec:Ki}.

In this scenario, the fidelity between the teleported state and the input state is given by  
$\mathcal{F}^K_\pm \coloneqq |\langle \phi|\rho_B^\pm|\phi\rangle|$,  
corresponding to the control being projected onto $\ket{0}\!\bra{0}$ and $\ket{1}\!\bra{1}$, respectively.
Since the input state $|\phi\rangle$ is a pure single-qubit state, it can be written as  
$|\phi\rangle \coloneqq a\ket{0} + \sqrt{1 - a^2} e^{i\eta} \ket{1}$, 
with $0 \leq a \leq 1$ and $0 \leq \eta < 2\pi$. Substituting this form of $|\phi\rangle$ along with $\rho_B^\pm$ into the expression for $\mathcal{F}^K_\pm$, we obtain that
\begin{align}
    \mathcal{F}_+^K &= \frac{2y(1-2a^2)^2 + 4a^2(1-a^2) + Xy(1-2a^2)^2}{2+Xy}\nonumber \\
    \mathcal{F}_-^K &= \frac{2y(1-2a^2)^2 + 4a^2(1-a^2) - Xy(1-2a^2)^2}{2-Xy}.
\end{align}

Let us define that $\alpha \coloneqq \cos(\theta_c/2)$ and $\beta \coloneqq \sin(\theta_c/2)e^{i\phi_c}$, where $\theta_c,\phi_c$ are the Bloch angles of the initial pure state of the control qubit with $0\leq \theta_c \leq {\pi}$ and $0\leq \phi_c < 2\pi$. Thus, we have $X \coloneqq \alpha\beta^* + \alpha^*\beta = \cos\theta_c \sin\phi_c$ with  $X\in[-1,1]$ and $y = p_1+p_2$ with $y\in[0,1]$. Therefore, the average fidelity in this case, computed by averaging over all pure single-qubit input states, is given by
\begin{align}
    \langle\mathcal{F}_\pm^K\rangle &\coloneqq \frac{2y\pm Xy+2}{3(2\pm Xy)}.
    \label{avg-fidelityK}
\end{align}

\begin{figure*}[!htb]
    \centering
    \includegraphics[width=0.80\linewidth]{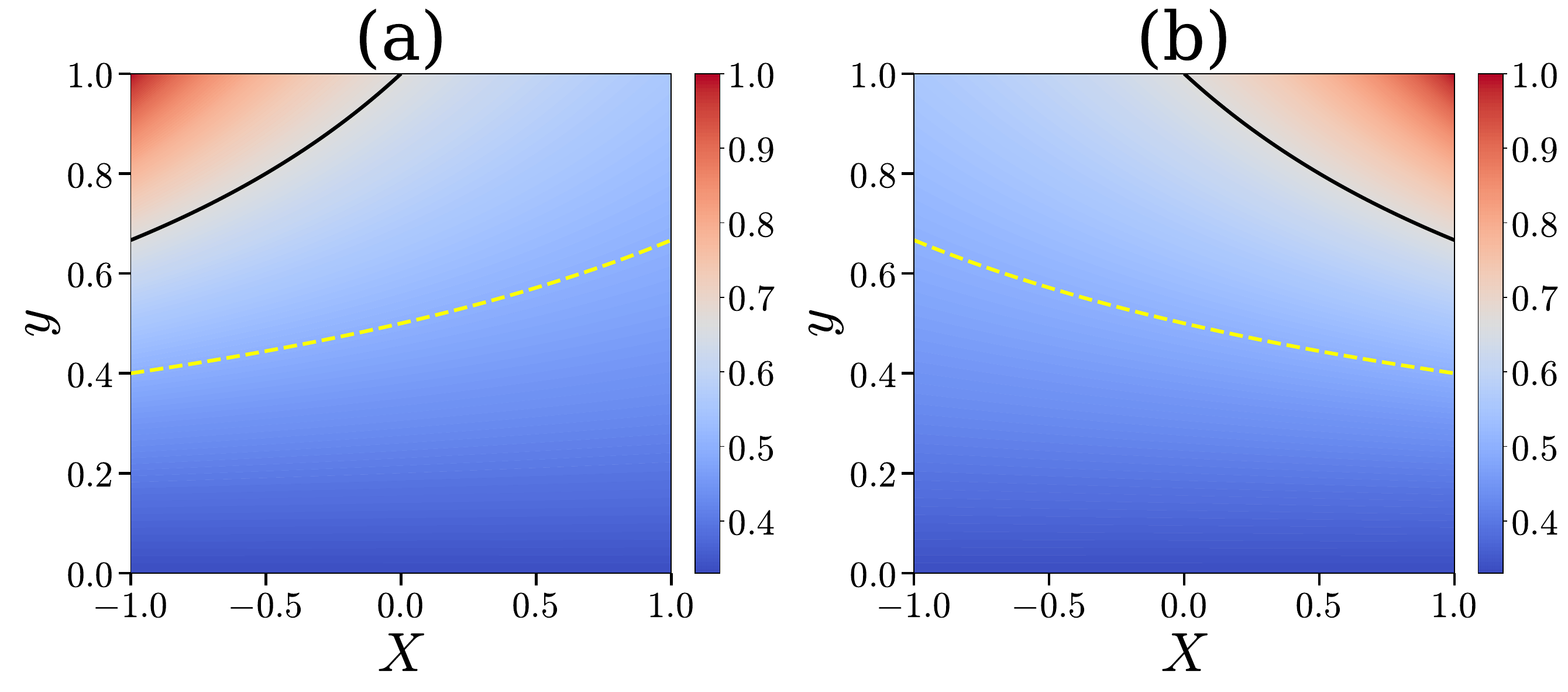}
   \caption{\textbf{\emph{Average teleportation fidelity when assisted by superposition of paths, with the shared state being diagonal in the computational basis.}} We present here the average fidelity using the standard teleportation protocol assisted by the superposition of paths, characterized by Kraus operators $\{M_{\mu,\nu}^K\}$, for shared separable states of the form given in Eq.~\eqref{state1}.
    We show the variation of (a) $\langle \mathcal{F}_+^K \rangle$ and (b) $\langle \mathcal{F}_-^K \rangle$, written in Eq.~\eqref{avg-fidelityK}, across the full range of parameters $X$ and $y$. $X = \alpha\beta^* + \alpha^*\beta$, where the control qubit is initially in the state $|\chi\rangle_C = \alpha\ket{0} + \beta\ket{1}$.
    The black solid line contour represents $\langle \mathcal{F}_\pm^K \rangle = \frac{2}{3}$, the classical average fidelity threshold, while the yellow dashed line marks $\langle \mathcal{F}_\pm^K \rangle = {1}/{2}$, corresponding to random guessing of the output state.
    The regions where the average fidelity exceeds ${2}/{3}$ appear in the upper left of (a) and the upper right of (b), indicate quantum advantage. These regions are bounded black solid line representing the condition $y(2 \mp X) = 2$. Thus, our protocol activates the quantum advantage in quantum teleportation even with shared separable states, enabled by the use of superposition of paths.}                         
    \label{fig-fidelityK}
\end{figure*}

We plot the average fidelity $\langle \mathcal{F}\pm^K \rangle$ as a function of $X$ and $y$ in Fig.~\ref{fig-fidelityK}, 
with $\langle \mathcal{F}+^K \rangle$ and $\langle \mathcal{F}_-^K \rangle$ plotted in panels (a) and (b), respectively.
In the Fig.~\ref{fig-fidelityK}, the black solid line marks the contour $\langle \mathcal{F}\pm^K \rangle = {2}/{3}$, while the yellow dashed line corresponds to $\langle \mathcal{F}_\pm^K \rangle = {1}/{2}$.
Recall that an average fidelity of ${2}/{3}$ represents the maximum average fidelity achievable using any classical standard teleportation protocol, whereas average fidelity of ${1}/{2}$ denotes the fidelity obtained by random guessing the output state. We can observe a region in the upper left corner of Fig.~\ref{fig-fidelityK}(a) and a region in the upper right corner of Fig.~\ref{fig-fidelityK}(b) where the average fidelity exceeds ${2}/{3}$.
Both these regions in Fig.~\ref{fig-fidelityK}(a) and Fig.~\ref{fig-fidelityK}(b) 
are bounded by the black solid line representing the condition $y(2 \mp X) = 2$, which serves as a lower bound on $y$; which implies that the protocol performs worse than the classical standard teleportation protocols below this threshold.
Thus, a quantum advantage can be achieved using the standard teleportation protocol assisted by the superposition of paths, characterized by the Kraus operators $\{M_{\mu,\nu}^K\}$, even when the shared separable state is of the form given in Eq.~\eqref{state1}.

Now, let us consider a second protocol where again the standard teleportation protocol is applied in conjunction with superposition of paths, but with a modified set of  Kraus operators $\{M^L_{\mu\nu}\}$.  The Kraus operators are given by  
\begin{align}
     M_{\mu \nu}^L = \ket{0}\langle0|\otimes\frac{1}{2}L_\mu  + \ket{1}\langle1|\otimes\frac{1}{2}L_\nu, 
     \label{map-L}
\end{align}
where the Kraus operators $\{L_i\}$ are written in Eq.~\eqref{phi-kraus}. Consequently, the average fidelity between the output state at Bob after applying the second protocol and the input state of Alice is found to be
\begin{align}
    \langle\mathcal{F}_\pm^L\rangle \coloneqq \frac{2z \pm Xz+2}{3(2\pm Xz)}.
\end{align}
 Here $z \coloneq p_0 + p_3 = 1-y$. More details are presented in Appendix~\ref{sec:Li}.

 {\color{black}
 In the scheme considered here, the two sets of Kraus operators, $\{K_\mu\}$ and $\{L_\mu\}$, are related by local unitary transformations. Consequently, the corresponding operators $\{M_{\mu\nu}^K\}$ and $\{M_{\mu\nu}^L\}$, which implement coherent superpositions of teleportation maps, are also locally unitarily equivalent.

Owing to this local-unitary equivalence, the teleportation protocols defined by $\{M_{\mu\nu}^L\}$ and $\{M_{\mu\nu}^K\}$ exhibit complementary behavior depending on the structure of the shared separable state. In particular, when $\{M_{\mu\nu}^K\}$ and $\{M_{\mu\nu}^L\}$ are applied to the separable states given in Eqs.~\eqref{type1} and \eqref{type2}, respectively., probabilistic unit-fidelity teleportation is achieved in both cases. 
% Conversely, when $\{M_{\mu\nu}^K\}$ and $\{M_{\mu\nu}^L\}$ are applied to the separable states in Eqs.~(13) and~(14), respectively, the optimal average fidelity reduces to $2/3$, which coincides with the classical benchmark.

The superposition of paths is implemented by the Kraus operators $M_{\mu\nu}^K$ ($M_{\mu\nu}^L$), by applying $K_\mu$ ($L_\mu$) in both of the interferometer arms. 
In the scheme considered here, the two sets of Kraus operators, $\{K_\mu\}$ and $\{L_\mu\}$, are related by local unitary transformations. Consequently, the corresponding operators $\{M_{\mu\nu}^K\}$ and $\{M_{\mu\nu}^L\}$, which implement coherent superpositions of teleportation maps, are also locally unitarily equivalent.
The control qubit $C$ provides an operational interpretation of the above Kraus operators. When $C$ is prepared in either $\ket{0}$ or $\ket{1}$ i.e., in the absence of coherence the sets $\{M_{\mu\nu}^K\}$ and $\{M_{\mu\nu}^L\}$ each reduce to the implementation of a single conventional teleportation map.
In contrast, when $C$ is initialized in a coherent superposition of $\ket{0}$ and $\ket{1}$, the operators $\{M_{\mu\nu}^K\}$ and $\{M_{\mu\nu}^L\}$ realize coherent superpositions of teleportation maps generated by the pairs $\{K_\mu\}$, $\{K_\nu\}$ and $\{L_\mu\}$, $\{L_\nu\}$, respectively.
}

\par
\begin{figure}
    \centering
    \includegraphics[width=0.8\linewidth]{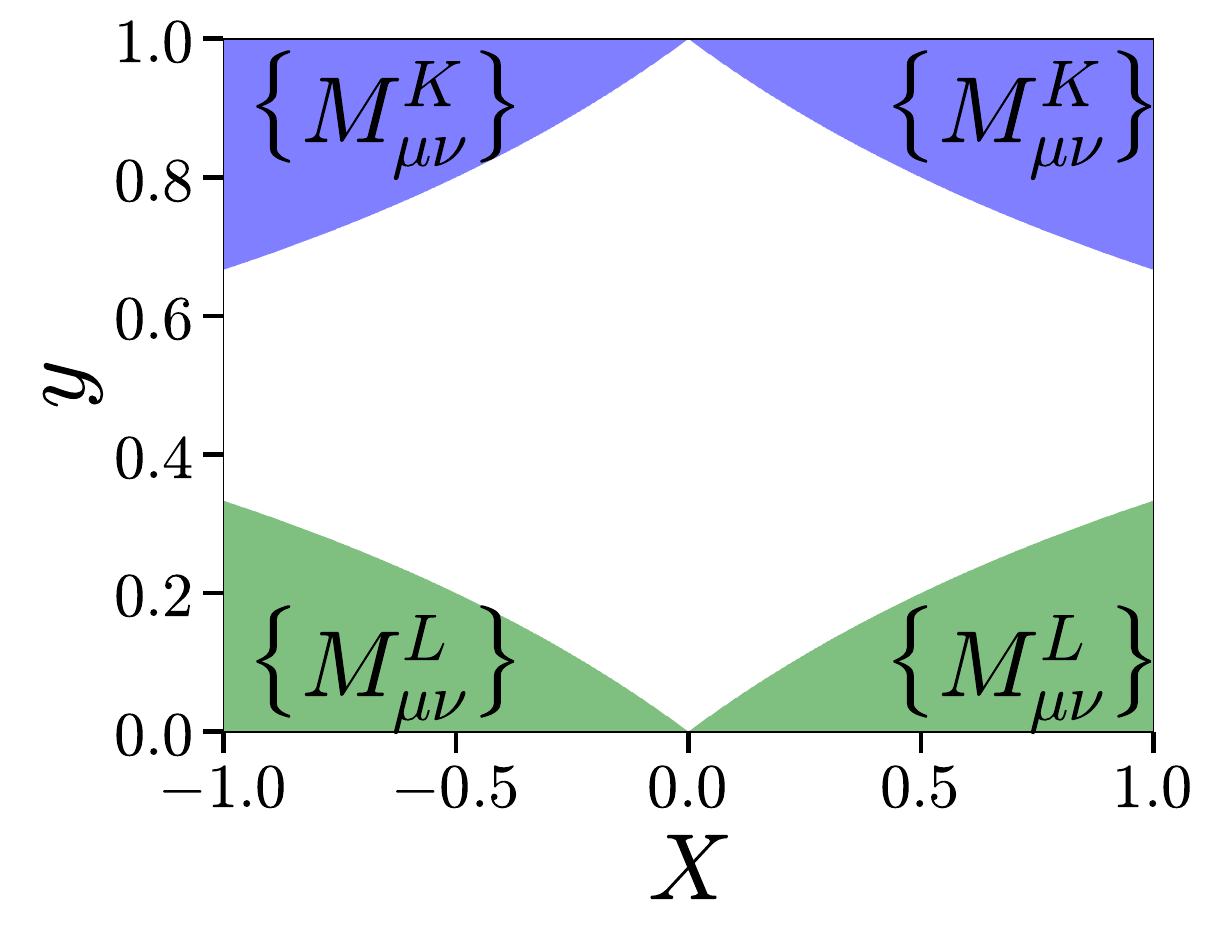}
  \caption{\textbf{\emph{\color{black}
  % Which Kraus operators provide quantum advantage in which region?
  Regions in parameter space with quantum advantage and the corresponding Kraus operators responsible.
  % for quantum advantage across distinct regions of the parameter space.
  }} We identify here the quantum advantage regions across the full range of $X$ and $y$ using our teleportation protocols, characterized by the Kraus operators $\{M_{\mu\nu}^K\}$ and $\{M_{\mu\nu}^L\}$, for shared separable states as given in Eq.~\eqref{state1}. The colored areas indicate parameter regimes where our  teleportation protocol outperforms classical protocols. Specifically, the blue-colored regions correspond to quantum advantage achieved using the protocol based on $\{M_{\mu\nu}^K\}$, while the green-colored regions indicate quantum advantage obtained via the protocol defined by $\{M^L_{\mu\nu}\}$.}
    \label{fig:Xy-schematic}
\end{figure}

\emph{\textbf{Quantum advantage in teleportation with coherent superposition of paths for shared separable states.}} For separable shared states $\rho_{A^\prime B}$, of the form as stated in Eq.~\eqref{state1}, it is possible to achieve quantum advantage probabilistically. When the shared state has a higher value of $y = p_1 + p_2$, that is, the state has a higher probability of being in the subspace spanned by $\{|01\rangle, |10\rangle\}$, then we are required to apply the map given by $M_{\mu\nu}^K$ as given in Eq.~\eqref{map-K}. Whereas, when $y$ is smaller and the state has a higher probability of being in the subspace spanned by $\{|00\rangle, |11\rangle\}$, then we are required to apply the map given by $M_{\mu\nu}^L$ as given in Eq.~\eqref{map-L}. 
More precisely, for larger values of $y$, satisfying $y(2\mp X)>2$, $M_{\mu\nu}^K$ provides quantum advantage, whereas, for smaller values of $y$ satisfying $(1-y)(2\mp X)>2$, $M_{\mu\nu}^L$ provides quantum advantage. It is shown in Fig.~\ref{fig:Xy-schematic}, where the colored region denotes quantum advantage with the corresponding protocol that is to be applied.   

Moreover, note that the teleportation fidelities  of either protocols,  $\langle \mathcal{F}_{\pm}^K\rangle$ or $\langle \mathcal{F}_{\pm}^L\rangle$ are independent of individual values of $p_i$ in Eq.~\eqref{state1}. Rather, they are dependent on the total probability of the subspace spanned by $\{|00\rangle, |11\rangle\}$ or  $\{|01\rangle, |10\rangle\}$. For intermediate values of $y$, we do not find any quantum advantage using coherent superposition of the discussed standard teleportation protocol, as shown in Fig.~\ref{fig:Xy-schematic}.   
\par
\emph{\textbf{Teleportation with unit fidleity is possible for separable states with coherent superposition of paths.}}
The teleportation fidelities $\langle \mathcal{F}_\pm^K\rangle$ and $\langle \mathcal{F}_\pm^L\rangle$ can take unit value for $y = 1$ and $y = 0$, respectively as shown in Fig.~\ref{fig-fidelityK}. 
From this we find that separable states of the form 
\begin{align}
    \rho_{A^\prime B} = p |00\rangle\langle 00| + (1-p) |11\rangle \langle 11|,
    \label{type2}
\end{align}
have $ y = 0$ with $0\leq p \leq 1$ and can  teleport any pure single qubit state with unit fidelity, probabilistically. For this, the control qubit is to be initialized either in $|+\rangle\coloneq (\ket{0} + \ket1)/\sqrt2$ (for which $X=1$)
or in $|-\rangle \coloneq (\ket0 - \ket1)/\sqrt2$ (for which $X  = -1$) and apply $M_{\mu\nu}^L$. If the control qubit $C$ was initialized in $|+\rangle$, then the state $|\phi\rangle$ is obtained when $\ket{1}$ is finally measured in the control qubit, whereas if $C$ was initialized in $|-\rangle$, when $\ket{0}$ is finally measured in $C$, we obtain the state with unit fidelity in $B$. In both the cases, the measurement outcome is obtained with a probability of $\frac{1}{4}$. 
\par Similarly, for the states of the form  \begin{align}
    \rho_{A^\prime B} = q |01\rangle\langle 01| + (1-q) |10\rangle \langle 10|,
    \label{type1}
\end{align}
we have $y = 1$ with $0\leq q \leq 1$, and $M_{\mu\nu}^K$ needs to be operated. For all values of $q$ unit fidelity is obtained with probability ${1}/{4}$ on measuring $C$ in computational basis and obtaining $\ket{0}$ when $C$ is initialized in $|-\rangle$ ($X = -1$) and on obtaining $\ket{1}$ when  initialized in $|+\rangle$ ($X = 1$). As a special case, we observe that for the extremum values of $p$ and $q$, the shared state $\rho_{A^\prime B}$ is a product state, still we achieve unit fidelity with probability ${1}/{4}$. 
% {\color{teal} For qubit teleportation, a protocol succeeding with probability $p$ achieves an average fidelity $\langle\mathcal{F}_p\rangle=(2f_p+1)/3$, where $f_p=\langle\Psi|\rho|\Psi\rangle$ is the singlet fraction of the effective shared state~\cite{horodecki1999}. Thus, to assess whether a success probability $p=1/4$ provides any advantage, one must examine the maximal singlet fraction attainable at that probability.

% For classical protocols restricted to separable shared states, the singlet fraction is bounded by $f_p\leq 1/2$, even allowing arbitrary post-selection and vanishingly small success probability~\cite{Massar1995,Kent1998}. Consequently, the average fidelity of any such protocol, including those post-selecting on only $1/4$ of the outcomes, is upper bounded by $2/3$.

% Therefore, post-selection alone cannot overcome the classical teleportation limit. In this sense, the unit-fidelity events occurring with probability $1/4$ in our protocol are genuinely nonclassical. Moreover, probabilistic teleportation schemes such as~\cite{Agrawal2002} also fail to provide any advantage when the shared resource is separable, confirming that entanglement is required to surpass the classical bound.
% }~\textcolor{red}{[report e ja likhechi otai thik korte hobe/amar sathe kotha bolte hobe. bki sob thik achhe.]}
{\color{black} 
% REDUCE the SIZE of THIS PARA.......
For qubit teleportation, suppose that the two parties share a bipartite state $\rho$. 
The protocol under consideration may be probabilistic; let $p$ denote the probability of successful implementation. 
A closely related dual question is how close one can transform $\rho$ to a maximally entangled singlet state with the same probability $p$. 
The performance of such a protocol is quantified by the singlet fraction of the effective (post-selected) shared state, defined as 
$F_p = \langle \Psi | \rho_p | \Psi \rangle$, where $|\Psi\rangle$ is a maximally entangled Bell state and $\rho_p$ denotes the normalized state conditioned on success. 
For qubits, the corresponding average teleportation fidelity is given by $\langle \mathcal{F}_p \rangle = \frac{2F_p + 1}{3}$, establishing a direct relation between the singlet fraction and teleportation performance~\cite{horodecki1999}.
Therefore, to assess whether a success probability $p = 1/4$ offers any genuine advantage, one must compare the maximum singlet fraction attainable at that probability. 
In a classical protocol where the shared state is restricted to be separable, it is known that the singlet fraction is bounded by $F_p\leq 1/2$~\cite{Massar1995,Kent1998}. 
% Importantly, this bound holds even if one allows arbitrary post-selection and arbitrarily small success probability. Consequently, the average fidelity of any classical protocol, including those post-selecting on only $1/4$ of the outcomes, is upper bounded by $2/3$.
{\color{black}Thus, post-selection alone does not allow classical strategies to surpass the classical teleportation limit. We give a general classical analysis in Appendix~\ref{final-app}.}
In this sense, the unit-fidelity events occurring with probability $1/4$ in our protocol represent a genuinely nonclassical feature. }

\emph{\textbf{Quantum teleportation with unit fidelity and a certain non-zero probability is possible for any shared pure product state.}} Looking at Eq.~\eqref{type2} and~\eqref{type1}, it is natural to wonder if the same is true for any shared states of the form 
\begin{align}
    \rho_{A^\prime B} = r |\omega \gamma\rangle\langle \omega \gamma| + (1-r) |\bar{\omega} \bar{\gamma}\rangle \langle \bar{\omega} \bar{\gamma}|,
    \label{typegen}
\end{align}
where we have $0\leq r \leq 1$, $\langle \omega  \ket{\bom}  = 0$ and $\langle \gamma \ket{\bga} = 0$. 
{\color{black}We can clearly see that applying local unitaries can take Eq.~\eqref{typegen} to the form of Eq.~\eqref{type2} or~\eqref{type1}(see~\footnote{{\color{black}We can clearly see that Eq.~\eqref{type2} can be obtained from Eq.~\eqref{type1} using a local unitary of $\mathbb{I}_2\otimes \sigma_x$. Thus, it is enough to show that Eq.~\eqref{typegen} can be obtained from Eq.~\eqref{type2} by applying a local unitary. The required unitary is of the form, $U = U_1 \otimes U_2$, with $U_1 = |0\rangle\langle \omega| + |1\rangle\langle \bar{\omega}|$ and  $U_2 = |0\rangle\langle {\gamma}| + |1\rangle\langle \bar{\gamma}|$ and setting $r = p$.}}).
} 
Therefore, teleportation of any arbitrary pure single qubit state is possible with the help of superposition of paths for shared states of the form written in Eq.~\eqref{typegen}. In Appendix~\ref{sec:Advantage pure product}, we present the details regarding such a protocol.
{\color{black}We would like to reiterate that the shared state in Eq.~\eqref{typegen}, when used in standard teleportation gives an average fidelity of ${2}/{3}$ as shown in Appendix~\ref{standard-tele-shared-sep}.} 
As a special case, note that any shared pure product state of the form $\rho_{A^\prime B} = |\omega \gamma \rangle \langle \omega \gamma|$,  can  teleport with unit-fidelity any pure state, with probability ${1}/{4}$, when the standard teleportation is applied with superposition of paths.
{\color{black} We extend our results to the teleportation of a general pure qudit. 
We find that, when only two paths are considered, each implementing the conventional qudit teleportation map, the protocol achieves an average fidelity exceeding the classical limit in arbitrary dimensions, even when the shared resource state is a pure product state (see Appendix~\ref{higer-dim-generalizations}). 
We leave the extension of this protocol to multiple coherent paths, as well as the investigation of whether alternative protocols could further enhance the teleportation fidelity, as directions for future work.}

\subsection{Teleportation in conjunction with superposition of indefinite causal order of maps for shared pure product states}
{\color{black}
In this subsection, we implement the superposition of indefinite causal order of maps in the standard teleportation protocol to teleport any pure single-qubit state for a product shared state. 
Till now, quantum teleportation using indefinite causal order of quantum maps has been studied considering noisy singlet state as shared state~\cite{Mukhopadhyay_2020, Cardoso-Isidoro_2020, Caleffi2020, Ban2022, dey2025}.
We here consider shared pure product states for the teleportation protocol.
For this, we use the theorem in Ref.~\cite{Bowen2001}, which states that the standard teleportation protocol is equivalent to a generalized depolarizing channel $\Lambda$, that is, if we want to teleport the state $\rho$, the output state under the action of the standard teleportation protocol is given by $\Lambda(\rho) = \sum_{i = 0}^3 p_i \sigma_i \rho\sigma_i$. 
We apply the superposition of indefinite causal order on the teleportation protocol. 
We obtain the final teleported state to be, $\rho_S' = \frac{1}{2}(|\phi\rangle\langle\phi| + \sigma_z |\phi\rangle\langle\phi|\sigma_z)$. See Appendix~\ref{teleportation-switch-app} for detailed derivation.
Therefore, the average fidelity over all pure states in this case will be ${2}/{3}$.  Thus, the standard quantum teleportation protocol with the superposition of causal order of quantum maps does not provide any quantum advantage over the standard teleportation protocol when the shared state is in pure product form, unlike standard teleportation with the superposition of paths discussed in the previous subsection.
}

\begin{figure}
    \centering
    \includegraphics[width=0.80\linewidth]{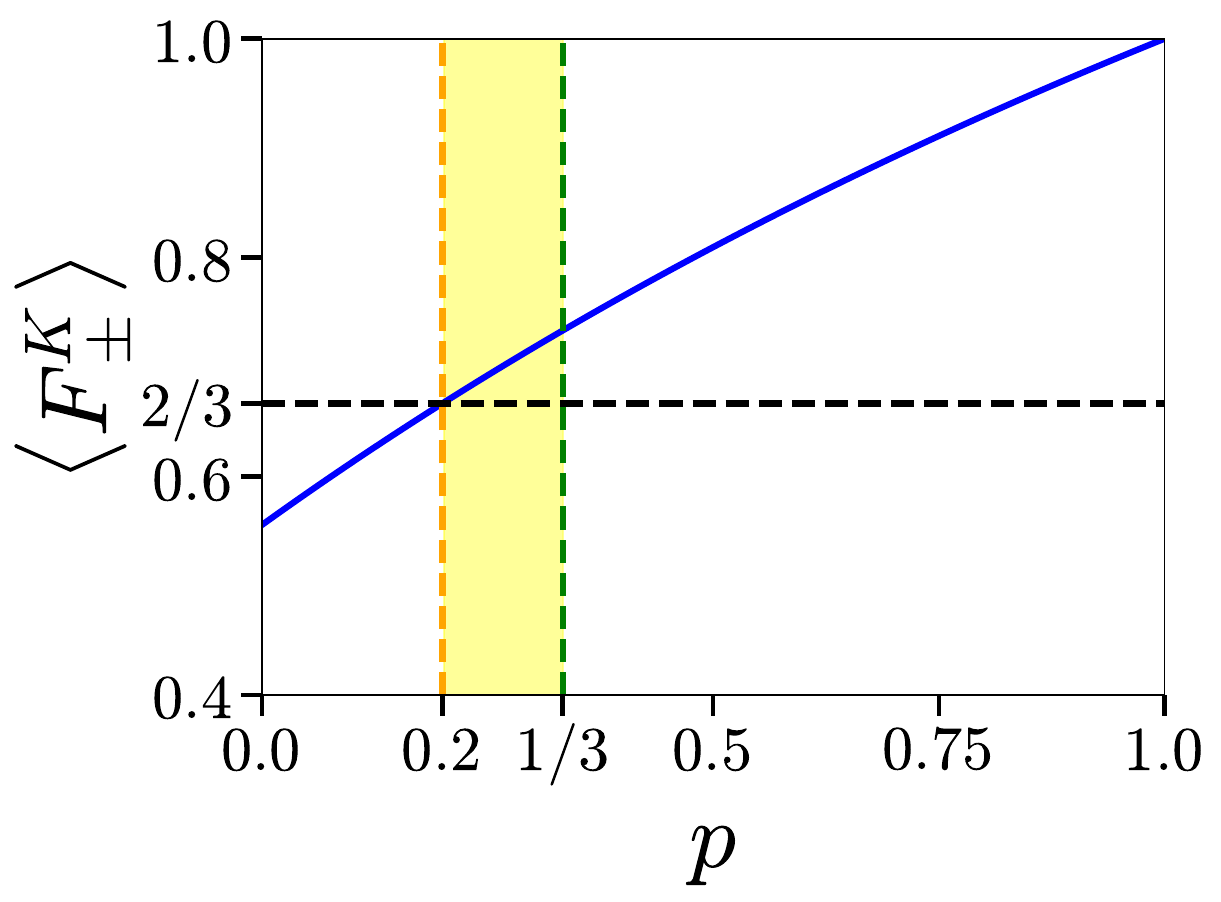}
   \caption{\textbf{\emph{Average fidelity for quantum teleportation assisted by path superposition for shared Werner states.}} We present here the average fidelity for different values of $p$ when standard teleportation is combined with a superposition of paths—characterized by the Kraus operators $\{M_{\mu\nu}^K\}$—for a shared two-qubit Werner state as defined in Eq.~\eqref{werner-state}. 
    The green dashed line marks $p = {1}/{3}$, below which the Werner state is separable. 
    The classical teleportation fidelity limit of ${2}/{3}$ is indicated by the black dashed line. The maximum average fidelity, $\langle F_\pm^K \rangle$, achieved using our protocol (based on the Kraus operators $\{M_{\mu\nu}^K\}$) is shown by the blue solid line, which is optimized over all parameters of the control qubit. 
    {\color{black} Conventional entanglement-assisted teleportation exhibits quantum advantage only for $p > {1}/{3}$, where the Werner state is entangled. In contrast, our protocol achieves $\langle \mathcal{F} \rangle > {2}/{3}$ already for $p > {1}/{5}$.
    The yellow shaded region, bounded below by $p = {1}/{5}$ (orange dashed line) and above by $p = {1}/{3}$ (green dashed line), highlights a range of separable Werner states that nonetheless exhibit quantum advantage when the standard teleportation protocol is applied in conjunction with a superposition of paths via the Kraus operators $\{M_{\mu\nu}^K\}$.}}
  \label{fig:werner}
\end{figure}
\begin{figure*}
    \centering
    \includegraphics[width = 0.8\linewidth]{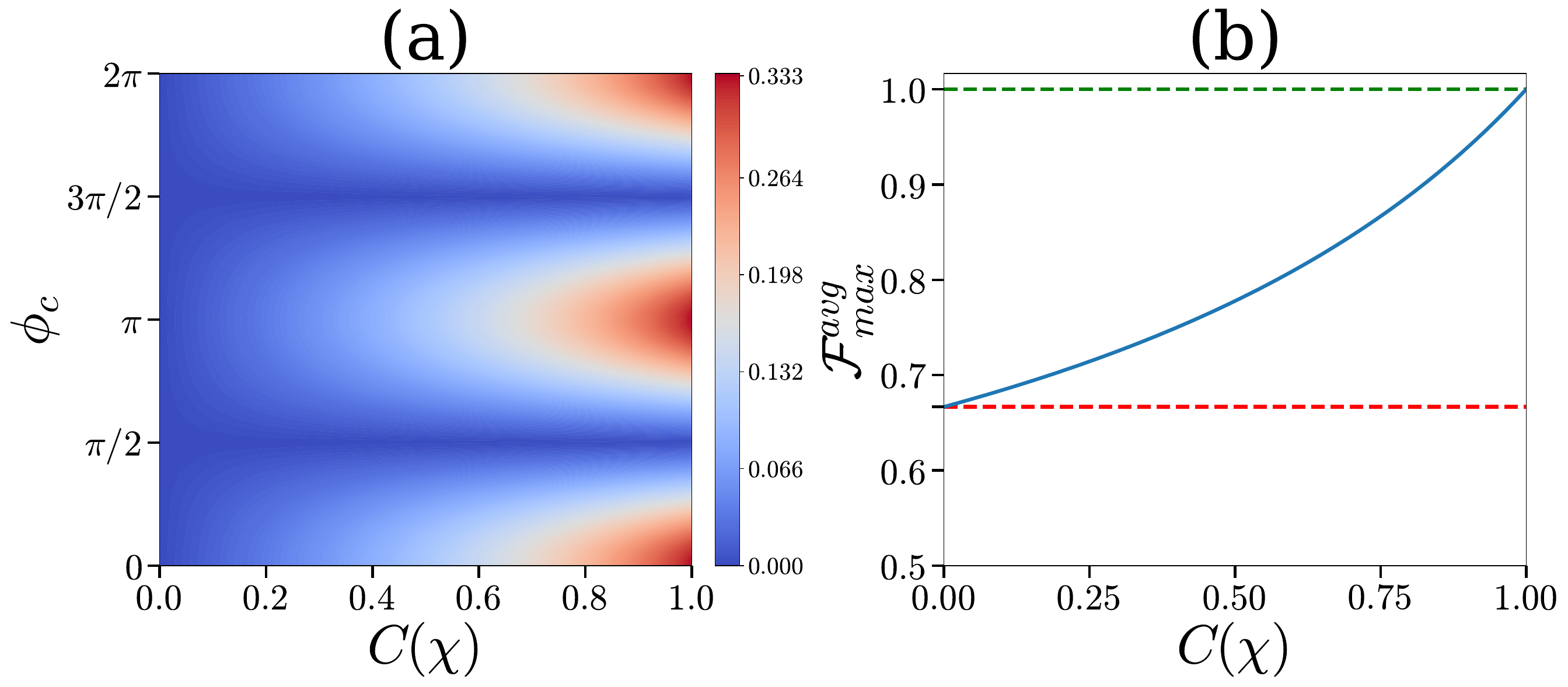}
    \caption{\textbf{\emph{Role of quantum coherence of the initial control qubit state in achieving quantum advantage using our teleportation protocol}}.  We plot the quantity $\mathcal{F}^K_{\text{adv}} \coloneq \mathcal{F}_{\text{max}}^{\text{avg}} - \frac{2}{3}$, where $\mathcal{F}_{\text{max}}^{\text{avg}} \coloneq \max\{\langle\mathcal{F}_{+}^{K}\rangle,\langle\mathcal{F}_-^K\rangle\}$ as presented in Eq.~\eqref{avg-fidelityK} with respect to $C(\chi)$ and $\phi_c$. 
    (a) We plot $\mathcal{F}^K_{\text{adv}}$, which denotes the quantum advantage on applying the Kraus operators $\{M^K_{\mu\nu}\}$ with the shared state being of the form of Eq.~\eqref{type1}. The Hadamard gate is applied on the control qubit  and it is consequently measured in the computational basis to obtain the state at Bob's place after the application of $\{M^K_{\mu\nu}\}$ on joint $CAA^\prime B$ state. We observe that when $\phi_c = \frac{\pi}{2},\frac{3\pi}{2}$,  $\mathcal{F}^K_{\text{adv}}$ becomes zero for any value of coherence of initial control qubit. 
    (b) Here all are same as (a), but only difference is that the unitary $U_{\xi, \zeta}$ is applied instead of the Hadamard gate on control qubit. For this case,  $\mathcal{F}^\text{avg}_{\text{max}}$ is plotted with blue solid line. The average fidelity  $\mathcal{F}_{\text{max}}^{\text{avg}}=\frac{2}{3}$ (shown with red dashed line), denotes the critical threshold line, any point above this line denotes quantum advantage. We observe that in this case the quantum advantage is restored as $\mathcal{F}_{\text{max}}^{\text{avg}} >\frac{2}{3}$ (as shown in Eq.~\eqref{avg-fid-U}) when the initial control qubit state has non-zero quantum coherence. Moreover, note that $\mathcal{F}_{\text{max}}^{\text{avg}}$ is a monotonic function of $C(\chi)$. At the maximum value of quantum coherence of initial control qubit, $\mathcal{F}_{\text{max}}^{\text{avg}} = 1.0$ (shown with green dashed line). Thus, quantum coherence is sufficient for achieving quantum advantage using our teleportation protocol.}
    \label{fig-coherence}
\end{figure*}

\subsection{Teleportation with superposition of paths for shared separable Werner states}
\label{sec:Werner-app}
Here, we analyze the quantum advantage achieved in the standard quantum teleportation protocol assisted by a superposition of paths—characterized by the Kraus operators $\{M_{\mu\nu}^K\}$—when the shared states are two-qubit Werner states.
Any two-qubit Werner states can be written as
\begin{align}
    \rho^W_{A^\prime B} = p\ket{\beta}\bra{\beta} + \frac{1-p}{4} \mathbb{I}_{4},
    \label{werner-state}
\end{align}
where $0\leq p \leq 1$ and $\ket{\beta}$ is one of the four Bell states. 
Let us illustrate how the superposition of maps can assist in quantum teleportation when the shared state is a two-qubit Werner state by looking at the case where  
$|\beta\rangle \coloneq |\psi^-\rangle$,

In this section, we apply the Kraus operators $\{M_{\mu\nu}^K\}$ that was introduced in the previous section to teleport $\ket{\phi}_A\bra{\phi}$ using shared state $\rho_{A^\prime B}^W$ (written in Eq.~\eqref{werner-state}) from $A$ to $B$. Our final state after applying $\{M_{\mu\nu}^K\}$ is of the same for as in Eq.~\eqref{eq:total_operation}, with the initial joint Alice-Bob state $\rho_{AA^\prime B}$ is of the form $\rho_{AA^\prime B}^i \coloneqq |\phi\rangle_A\langle\phi|\otimes \rho^W_{A^\prime B}$.
We find that
\begin{align*}
    \sum_{\mu = 0}^3 K_\mu \rho_{AA^\prime B}^i K_\mu^\dagger &= \psi^-_{AA^\prime}\otimes\left(p|\phi\rangle\langle\phi| + \frac{1-p}{2}\mathbb{I}_{2}\right)_B, \\
    \sum_{\mu,\nu = 0}^3 K_\mu \rho_{AA^\prime B}^i K_\nu^\dagger &= \psi^-_{AA^\prime}\otimes\left[(1-p)\sigma_z|\phi\rangle\langle\phi|\sigma_z\right]_B.
\end{align*}
After applying the Hadamard gate on the control qubit $C$ of $\rho_{AA^\prime B}^f \coloneqq \sum_{\mu,\nu} M_{\mu,\nu}^K \rho_{AA^\prime B}^i (M_{\mu,\nu}^K)^\dagger$ and measuring it in the $\{\ket{0},\ket{1}\}$ basis, we obtain the state at Bob
\begin{align}
\rho_B^{\pm} &= \frac{4\rho^p_{\text{dep}} \pm X(1-p)\sigma_z|\phi\rangle\langle\phi|\sigma_z }{4\pm X(1-p)}, \nonumber \\
\text{with} \hspace{5mm} p_{\pm} &= \frac{4\pm X(1-p)}{8},
\end{align}
where $\rho^p_{\text{dep}} \coloneq p|\phi\rangle\langle\phi| + \frac{1-p}{2}\mathbb{I}_{2}$.
As defined earlier $X \coloneq \alpha\beta^* + \alpha^*\beta$. Subsequently, the fidelity $F_{\pm}^K$ for shared two-qubit Werner states is given by
\begin{align*}
    F^K_{\pm} = \frac{2(1+p) \pm X(1-p)(1-2a^2)^2}{4\pm X(1-p)},
\end{align*}
where, the state to be teleported  $\ket{\phi} = a\ket{0} + \sqrt{1-a^2}e^{i\eta}\ket{1}$ with $0\leq a\leq 1$ and $0\leq \eta <2\pi$. Thus, the average fidelity in this case comes out to be
\begin{align}
    \langle F_\pm^K \rangle = \frac{6(1+p) \pm X(1-p)}{12 \pm 3X(1-p)}. 
    \label{fidelity-werner}
\end{align}
We plot the average fidelity, $\langle F_\pm^K \rangle$ with respect to $p$ at $X = \mp1$, in Fig.~\ref{fig:werner}. We know that the average fidelity exceeds $\frac{2}{3}$ when the two-qubit Werner state written in Eq.~\eqref{werner-state} is entangled, that is $p > \frac{1}{3}$ in the standard teleportation protocol.  
Despite the fact that the two-qubit Werner states are separable for $\frac{1}{5} < p < \frac{1}{3}$, quantum advantage is observed even for $\frac{1}{5} < p < \frac{1}{3}$ because the average fidelity exceeds $\frac{2}{3}$ in this range of $p$, by applying superposition of maps in standard teleportation protocol. 
{\color{black} The yellow-shaded region in Fig.~\ref{fig:werner} highlights the parameter interval $\frac{1}{5} < p < \frac{1}{3}$. In this regime, the Werner state is separable and conventional entanglement-assisted teleportation cannot exceed the classical fidelity threshold $\frac{2}{3}$. Remarkably, our protocol achieves $\langle \mathcal{F} \rangle > \frac{2}{3}$ already for $p > \frac{1}{5}$, thereby exhibiting a quantum advantage in a parameter region where conventional teleportation fails. 
{The shaded region therefore represents an additional operational regime, beyond standard teleportation, in which quantum advantage is enabled 
% by process entanglement 
despite the absence of shared state entanglement.}
}
% This quantum advantage is shown by yellow shaded region in Fig.~\ref{fig:werner}. 
Although we have proved our results here for Werner states constructed by mixing $\psi^-$ with $\mathbb{I}_4$, the results will hold for Werner states constructed from other Bell states as well.
Thus, quantum advantage in quantum teleportation can be activated by applying superposition of paths in standard teleportation protocol even for a class of shared separable two-qubit Werner states with $\frac{1}{5} < p < \frac{1}{3}$.

{\color{black}
\section{Resources required for quantum advantage in teleportation using superposition of paths}

Quantum resources are non-classical properties of quantum systems that enable tasks that are impossible or inefficient within classical physics. Quantum resource theories~\cite{ChitamberRMP2019,Gour_2025} provide a general framework to characterize and quantify such resources. Several nonclassical properties of quantum systems, including entanglement~\cite{ent-re-1,ent-re-2,ent-re-3,ent-re-4,ent-re-5}, purity~\cite{thermo-re-2,purity-re-2}, quantum coherence~\cite{coherence-re-1, Baumgratz2014,coherence-re-3}, asymmetry~\cite{asymmetry-re-1,asymmetry-re-2,asymmetry-re-3}, magic~\cite{magic-re-1,magic-re-2,magic-re-3}, and athermality~\cite{thermo-re-1,thermo-re-2,thermo-re-3,thermo-re-4,thermo-re-5,thermo-re-6,thermo-re-7,thermo-re-8}, have been identified as quantum resources for a wide variety of quantum information-processing tasks.
%In this framework, resources are defined relative to a restricted set of free states and free operations, thereby allowing one to study their manipulation, conversion, and operational significance in a rigorous manner. Resource-theoretic frameworks have been extensively developed for several quantum resources, including entanglement~\cite{RevModPhys.81.865} and coherence~\cite{Strelsov2017}. 

In this section, we first demonstrate that quantum coherence is both necessary and sufficient for obtaining quantum advantage in teleportation using superposition of paths. We subsequently discuss process entanglement as an operational resource underlying the probabilistic teleportation protocol enabled by superposition of paths.}

\subsection{Role of coherence in quantum advantage of teleportation }
\label{sec:role-coherence}
In this section, we investigate the role of quantum coherence of the initial control qubit in achieving quantum advantage {\color{black} using our discussed quantum teleportation protocol for the shared state of the form Eq.~\eqref{typegen}.
Thus, we focus on the extremal cases $y = 0$ or $y = 1$ [of Eq.~\eqref{state1}]. In this way, the protocol operates at the optimal boundary permitted under LOCC operations.}
From section~\ref{sec3}, it is very clear that if the control qubit $C$ is initialized in a mixed state $\rho_C \coloneqq |\alpha|^2 \ket{0}\langle 0| + |\beta|^2 \ket{1}\langle 1|$ with $|\alpha|^2 +|\beta|^2 = 1$, the quantum advantage that we observed for even separable shared states vanishes. This is because the off-diagonal terms in Eq.~\eqref{eq:total_operation} turns out to be zero in this case. {\color{black} These initial states of the control qubit that are diagonal in computational basis have no quantum coherence.}
% , hence are not useful in the discussed teleportation protocol.} 
% {\color{black} In the scheme introduced in the current work, coherence in the initial control qubit state leads to extra terms as seen in Eq.~\eqref{eq:total_operation}. If the control qubit has $\gamma$ in one of the off-diagonal entries and $\gamma \neq 0$, then the initial coherence is non-zero. This provides a  quantum advantage. In our analysis we defined $X$ which quantifies this coherence. In this case, $X = \gamma + \gamma^*$. In Eq.~\eqref{avg-fidelityK}, we present how the average fidelity depends on $X$.}
% {\color{black}Since $\rho_C$ is diagonal in the computational basis, its coherence vanishes and it 
{\color{black} They cannot generate a coherent superposition of paths, which is essential for the teleportation protocol considered here, hence are not useful in the discussed teleportation protocol.

For pure control states with nonzero coherence $C(\chi)$, Eq.~\eqref{avg-fid-U} explicitly relates coherence to the probabilistic maximum average teleportation fidelity, showing that increased coherence directly enhances the quantum advantage.

For mixed control states, $\rho_C$ can be decomposed into a convex combination of a coherence-free state and a coherent pure state~\cite{Coh12,Coh23}. In this case, path superposition occurs only in the coherent fraction of runs, and the amount of coherence bounds the attainable average fidelity. Thus, coherence of the control qubit is the key resource governing the protocol’s performance.
}
It is therefore, very important to study the relation between coherence and the quantum advantage that we obtain.
\par
Quantum coherence of any quantum state, let's say $\rho$, is defined using the $l_1$-norm~\cite{aberg2006quantifying, Baumgratz2014,Strelsov2017, hu2018} as follows: 
\begin{align*}
    C(\rho) \coloneq \sum_{i,j,i\neq j} |\langle i|\rho|j\rangle |,
\end{align*}
where, $\{|i\rangle\}$ denotes a basis. Consequently, $C(\rho)$ is a basis-dependent quantity. 

It can be clearly seen from Eq.~\eqref{eq:total_operation}, the quantum advantage in our discussed protocol appears from the terms that are off-diagonal in the part of the control qubit $C$. If these parts were zero, no quantum advantage would have been obtained. Hence, we conclude that coherence of the initial control qubit state is necessary for observing quantum advantage with shared separable state in our discussed teleportation protocol.
\par 
{Let us now investigate whether quantum coherence is sufficient for observing quantum advantage in the discussed protocol.} Quantum coherence of the initial control qubit state $\chi \coloneqq |\chi\rangle_C\langle\chi|$ will be 
\begin{align}
   C(\chi) = |\alpha\beta^*| + |\alpha^*\beta| = \sin\theta_c , 
\end{align}
where $\ket{\chi}_C \coloneqq \cos\frac{\theta_c}{2}\ket{0} + e^{i\phi_c}\sin\frac{\theta_c}{2}\ket{1}$ with $0\leq \theta_c \leq {\pi}$ and  $0\leq \phi_c < 2\pi$.
We realize that $X = C(\chi)\cos\phi_c$. In Fig.~\ref{fig-coherence}, we plot the quantity $\mathcal{F}^K_{\text{adv}} \coloneq \mathcal{F}_{\text{max}}^{\text{avg}} - \frac{2}{3}$ with respect to $\phi_c$ and $C(\chi)$, where $\mathcal{F}_{\text{max}}^{\text{avg}} \coloneqq \max(\langle\mathcal{F}_{+}^{K}\rangle,\langle\mathcal{F}_-^K\rangle)$, with $\langle\mathcal{F}_\pm^K\rangle$ is written in Eq.~\eqref{avg-fidelityK}. $\langle\mathcal{F}_\pm^K\rangle$ denotes the average fidelity obtained when the shared state is of the form of Eq.~\eqref{type1} and the Kraus operators $\{M^K_{\mu\nu}\}$ are applied. 
$\mathcal{F}^K_{\text{adv}}$ quantifies the amount of quantum advantage in our protocol as illustrated in Fig.~\ref{fig-coherence}(a). Note that despite having $C(\chi) = 1$, we have no quantum advantage, for some specific values of $\phi_c$ for example when $\phi_c = \frac{\pi}{2}, \frac{3\pi}{2}$. For instance, consider the initial state control qubit state of $|\chi\rangle_c \coloneqq \frac{1}{\sqrt{2}}\ket{0} \pm i\frac{1}{\sqrt{2}}\ket{1}$, see that the control qubit state $|\chi\rangle_c$ has maximum quantum coherence when $l_1$-norm is used as a measure of quantum coherence but provides no quantum advantage in teleportation of any quantum state. 

To resolve this, instead of the Hadamard gate, we need to apply some other unitary on the control qubit $C$ before measuring it in computational basis. For this, let us consider a generic unitary operator on the the control qubit $C$ of the form 
\begin{align}
    U_{\xi, \zeta} \coloneqq \frac{1}{\sqrt 2}
    \begin{pmatrix}
e^{i\xi} & -e^{-i\zeta}  \\
e^{i\zeta} & e^{-i\xi} 
\end{pmatrix},
\label{Unitary_coh}
\end{align}
with $0\leq \zeta, \xi <2\pi$. After applying $U_{\xi, \zeta}$ on control qubit, we measure the control qubit in the computational basis, $\{\ket{0}, \ket{1} \}$. This is equivalent to measuring in the basis $\{|+;\zeta,\xi\rangle , |-;\zeta,\xi\rangle\}$ which are defined in Eq.~\eqref{U-on01}. 
Applying $U_{\xi,\zeta}$ from Eq.~\eqref{Unitary_coh} on the state post applying  the coherent superposition of the standard teleportation protocol, as in Eq.~\eqref{eq:total_operation}, we get 
\begin{widetext}
\begin{align*}
    \rho'_{CAA^\prime B} \coloneqq& |\alpha|^2 U_{\xi,\zeta}\ket{0}_C\langle0|U_{\xi,\zeta}^\dagger\otimes\psi^-_{AA^\prime} \otimes \rho_d + \frac{\alpha\beta^*}{2} U_{\xi,\zeta}\ket{0}_C\langle1|U_{\xi,\zeta}^\dagger\otimes\psi^-_{AA^\prime} \otimes \sigma_z|\phi\rangle\langle\phi|\sigma_z \nonumber\\
    &+  \frac{\alpha^*\beta}{2} U_{\xi,\zeta}\ket{1}_C\langle0|U_{\xi,\zeta}^\dagger\otimes\psi^-_{AA^\prime} \otimes \sigma_z|\phi\rangle\langle\phi|\sigma_z + |\beta|^2 U_{\xi,\zeta}\ket{1}_C\langle1|U_{\xi,\zeta}^\dagger\otimes\psi^-_{AA^\prime} \otimes \rho_d,
% \label{rho_tot_full}
\end{align*}
\end{widetext}
with $\rho_d = \frac{1}{2}(|\phi\rangle\langle\phi| + \sigma_z|\phi\rangle\langle\phi|\sigma_z) $. Note that 
\begin{align}
    |+; \zeta, \xi\rangle \coloneqq U_{\xi,\zeta}\ket{0} &= \frac{1}{\sqrt 2} (e^{i\xi}\ket{0} + e^{i\zeta}\ket{1}), \nonumber\\
    |-; \zeta, \xi\rangle \coloneqq U_{\xi,\zeta}\ket{1} &= \frac{1}{\sqrt 2} (-e^{-i\zeta}\ket{0} + e^{-i\xi}\ket{1}).
    \label{U-on01}
\end{align}
After this, measuring the control qubit of $\rho'_{CAA^\prime B}$ in the $\{\ket{0}, \ket{1}\}$ basis, the state received at Bob is given by 
\begin{align*}
    \rho^\pm_B &= \frac{2\rho_d  \mp C(\chi)\cos(\phi_c - \xi- \zeta) \sigma_z|\phi\rangle\langle\phi|\sigma_z}{2 \mp C(\chi)\cos(\phi_c - \xi- \zeta)}, \\
   \text{with} \hspace{4mm} p^\pm &=  \frac{2 \mp C(\chi)\cos(\phi_c - \xi- \zeta)}{4}.
\end{align*}
We choose the Unitary $U_{\xi,\zeta}$ such that $\xi + \zeta = \phi_c$. The maximum of the average fidelity optimizing over all control qubit's parameters is given by
\begin{align}   \mathcal{F}_{\text{max}}^{\text{avg}} = \frac{4 - C(\chi)}{6 - 3C(\chi)} \geq \frac{2}{3}.
    \label{avg-fid-U}
\end{align}
{\color{black} Although the derivation of Eq.~\eqref{avg-fid-U} was carried out assuming an initially pure control qubit state, the same procedure can be straightforwardly extended to the case of an initially mixed control qubit state. }
Thus, for non-zero coherence of the control qubit, we can always achieve quantum advantage in terms of average fidelity of the teleported state, using our protocol with the shared state being of the form Eq.~\eqref{type2}. This is illustrated in Fig.~\ref{fig-coherence}(b). Here, we analyzed the case for $\langle \mathcal{F}_{\pm}^K\rangle$, but the same arguments hold true for  $\langle \mathcal{F}_{\pm}^L\rangle$ with shared state of the form of Eq.~\eqref{type1}. 
{\color{black}In the absence of coherence in the control qubit ($C(X)\to 0$), the superposition of paths protocol reduces to a classical probabilistic mixture of the two paths. Since both paths implement the same teleportation map, the protocol becomes equivalent to standard teleportation without entanglement. For separable shared states of the form in Eqs.~\eqref{type2} or \eqref{type1}, the achievable average fidelity is $\mathcal{F}_{\max}^{\mathrm{avg}}=2/3$, coinciding with the classical benchmark.
Introducing coherence in the control qubit enables a genuine superposition of paths and activates a quantum advantage: any nonzero coherence suffices to surpass the classical limit. In the maximally coherent case ($C(X)\to 1$), the protocol attains unit teleportation fidelity with success probability $1/4$. Thus, Fig.~\ref{fig-coherence}(b) identifies coherence as the resource that lifts the protocol beyond optimal classical performance for these shared states.
}
{\color{black}
% Fig.~\ref{fig-coherence}(b) shows that any nonzero coherence in the control qubit enables a quantum advantage even when the shared state is separable, while strictly zero coherence forbids it. For a fully incoherent control (diagonal in the path basis), the protocol reduces to a classical mixture of LOCC maps and the average teleportation fidelity remains bounded by the classical limit. 
Thus, coherence is a necessary enabling resource, with its magnitude determining the size of the advantage; there is no finite coherence threshold, but zero coherence strictly precludes any advantage.

Although arbitrarily small coherence suffices in principle, the resulting advantage scales with the off-diagonal elements of the control state and becomes vanishingly small for weak coherence. Finite visibility and decoherence therefore impose practical lower bounds on detectability. While preparing path superpositions is routine in photonic platforms, preserving coherence throughout conditional LOCC operations remains a nontrivial experimental challenge.}

{\color{black}
Experiments demonstrating superpositions of noisy channels in photonic systems have been reported in Ref.~\cite{Rubino2021}, and similar setups could, in principle, be adapted to implement the present teleportation scheme. However, realizing the operators $M_{\mu\nu}^K$ requires implementing the same Kraus operators $\{K_\mu\}$ coherently in both arms, where $\{K_\mu\}$ correspond to measurement and outcome-dependent unitaries on a distant party. Such coherent control of spatially separated LOCC operations has not yet been experimentally demonstrated. Our proposal should therefore be viewed as an operationally well-defined but currently unrealized extension of existing controlled-channel experiments. Notably, all required ingredients, Bell-state measurements, applying local unitary based on classical communication, and coherent superposition of channels, have been demonstrated independently.

The protocol discussed here, requires no shared entanglement, but uses classical communication as in standard teleportation. The only additional resource is a local control qubit enabling the superposition of paths. In this sense, the protocol trades shared entanglement for coherence of the control qubit.}

Hence, quantum coherence of the initial state of the control qubit, is both necessary and sufficient to achieve quantum advantage in teleportation using our teleportation protocol assisted with superposition of paths for the shared states of the form Eq.~\eqref{typegen}.

{\color{black}
\subsection{Process entanglement}
In the context of the present work, we use the term ``process entanglement" to refer to the non-classical resource emerging from the coherent superposition of quantum channels enabled by coherent control over distinct channels. More specifically, we identify process entanglement as the operational resource responsible for the observed quantum advantage in teleportation, which arises only when the following ingredients are simultaneously present:
(i) classical control over distinct paths,
(ii) quantum coherence in the control qubit, and
(iii) a genuine operational quantum advantage over all corresponding classically controlled or incoherent implementations.

The necessity of the first two ingredients is evident from the operational structure of the protocol. Without classical control of the channels, the superposition of alternative channel paths cannot be realized. Conversely, our results in Sec.~\ref{sec:role-coherence} explicitly demonstrate that the coherence of the control qubit is essential for obtaining the teleportation advantage; in its absence, the enhancement disappears. However, the mere presence of classical control and coherence alone is not sufficient for what we refer to as process entanglement. We reserve this terminology specifically for situations in which these ingredients collectively yield a demonstrable operational quantum advantage in the underlying information-processing task.

Furthermore, the process advantage vanishes when the Kraus operators associated with the two channels mutually commute, indicating that the mere presence of classical control and quantum coherence alone is not sufficient for what we refer to as process entanglement.

Thus, process entanglement is not defined solely by the structural presence of coherent control over channel superpositions, but by its operational utility in enabling a genuine quantum advantage in teleportation over all corresponding classically controlled or incoherent strategies.
%Furthermore, the process advantage vanishes when the Kraus operators associated with the two channels mutually commute, indicating that the mere presence of classical control and quantum coherence alone is not sufficient for what we refer to as process entanglement.

Since process entanglement is fundamentally an operation-based resource, the quantum processes that do not exhibit process entanglement will be regarded as free operations. In the context of the present work, these correspond to processes involving 
superposition of paths generated by coherent control over distinct channels but yielding no genuine quantum advantage in the teleportation task. 
For example, within such a framework, several measures can be defined to quantify the amount of process entanglement generated by a given quantum process. One natural possibility is a distance-based approach. For example, the amount of process entanglement associated with a quantum process can be defined as
% We agree that these observations do not yet constitute a complete resource-theoretic characterization of process entanglement. Developing such a framework, including the identification of free processes, free super-operations, monotones, and operational tasks, is an interesting and important direction for future investigation, but lies beyond the scope of the present work.
% Nevertheless, one may formally quantify the amount of process entanglement through a distance-based approach. For example, one may define
\begin{equation}
\mathcal{E}_{\mathrm{process}}(\Lambda_{\mathrm{tele}})
=
\min_{\Gamma \in \mathbb{S}_{\mathrm{tele}}}
\mathcal{D}\!\left(\Lambda_{\mathrm{tele}},\Gamma\right),
\end{equation}
where $\Lambda_{\mathrm{tele}}$ denotes an effective teleportation map which involves superposition of paths, $\mathbb{S}_{\mathrm{tele}}$ denotes the set of maps which involve superposition of paths and do not exhibit any quantum advantage in the teleportation task, and $\mathcal{D}$ is a suitable distance measure between quantum channels. 
Possible choices for $\mathcal{D}$ include the diamond norm distance, trace-norm distance between Choi states, or other channel distance measures~\cite{arXiv9806029,PhysRevA.71.062310, PhysRevA.97.012332}.
The diamond norm quantifies the optimal operational distinguishability between two channels through a supremum over all possible input states, including those entangled with ancillary systems. Therefore,  $\mathcal{D}_{\diamond}(\mathcal{E},\mathcal{F}) = \max_{\rho} ||(\mathbb{I}\otimes\mathcal{E})\rho - (\mathbb{I}\otimes\mathcal{F})\rho||_1$, with $||X||_1 = \text{Tr}\sqrt{X^\dagger X}$ is the trace-norm and $\rho$ is a quantum state on an extended Hilbert space. 
On the other hand, the trace-norm distance between Choi states measures the distinguishability of the corresponding Choi representations of the channels, i.e., $\mathcal{D}_\text{choi}(\mathcal{E},\mathcal{F}) = ||\rho_\mathcal{E} - \rho_{\mathcal{F}}||_1$, where $\rho_{\mathcal{M}}$ is the Choi state of the map $\mathcal{M}$. 
In this way we can define the process entanglement as an operation resource useful in activating teleportation using superposition of paths.
% Possible choices for $\mathcal{D}$ include the diamond norm distance, trace-norm,  distance between Choi states, or other channel distance measures~\cite{PhysRevA.71.062310, PhysRevA.97.012332}. 
}
\section{Conclusion}
\label{conclu}

{\color{black}
In this work, we showed that a coherent superposition of paths fundamentally modifies the resource requirements for quantum teleportation. By embedding the conventional teleportation protocol within a superposition of LOCC maps, we demonstrated that quantum advantage is achievable even when the shared state between sender and receiver is separable. In particular, separable states including pure product states can enable unit-fidelity teleportation with success probability $1/4$ when combined with path superposition, which is impossible within the standard framework. We further identified classes of separable Werner states that exhibit a similar advantage. Importantly, this enhancement does not arise from indefinite causal order; for pure product shared states, indefinite causal order alone provides no advantage.

Our results do not imply that coherence generically simulates entanglement. 
% Entanglement between sender and receiver remains absent throughout. 
Rather, coherence in control qubit induces interference between distinct LOCC teleportation maps, generating an effective form of process-level entanglement that enables probabilistic unit-fidelity teleportation. We showed that coherence of the control qubit is necessary and sufficient for this advantage: in its absence, the protocol reduces to a classical mixture of LOCC maps and the advantage vanishes.

Although we focused on qubits, the mechanism extends to higher-dimensional systems by replacing each qubit teleportation map with its higher-dimensional analogue. 
% {\color{blue}Notably, unlike standard entanglement-assisted teleportation, deterministic unit-fidelity teleportation cannot be achieved using process entanglement alone; whether this is possible remains an open question.}
{Notably, whether deterministic unit-fidelity teleportation can be achieved using superposition of paths with separable shared state, remains an open question.}
While our analysis applies to finite-dimensional systems, continuous-variable (CV) systems differ fundamentally. In infinite dimensions, maximally entangled states are unphysical (requiring infinite squeezing), and practical CV teleportation instead relies on finitely squeezed states that introduce unavoidable Gaussian noise. Moreover, several tools used in the finite-dimensional setting do not directly generalize. Extending our framework to CV systems is therefore nontrivial and remains an open problem. 

Although a coherent superposition of full teleportation protocols has not yet been realized experimentally, all required components have been demonstrated independently. Our results thus provide an operational framework for teleportation beyond shared entanglement.
}

% \section*{Acknowledgements}
% PG acknowledges support from the ‘INFOSYS scholarship for senior students’ at Harish-Chandra Research
% Institute, India.

% \vspace{-2mm}
\section*{Appendix}
\appendix
\setcounter{figure}{0}
\renewcommand{\thefigure}{A \Roman{figure}}

\section{\texorpdfstring{Standard teleportation protocol in conjunction with superposition of paths for shared states of the form in Eq.~\eqref{state1}}{Standard teleportation protocol with path superposition}}
Here, we present the detailed calculations of the standard teleportation protocol assisted by the superposition of paths, described by Kraus operators $\{M_{\mu\nu}^K\}$ and $\{M_{\mu\nu}^L\}$, for shared states of the form given in Eq.~\eqref{state1}.

\subsection{\texorpdfstring{Protocol characterized by Kraus operators $\{M_{\mu\nu}^K\}$}{Protocol characterized by Kraus operators M}}
\label{sec:Ki}
On applying the Kraus operators $\{M_{\mu \nu}^K\}$ on the joint state in control-Alice-Bob, $\rho_{CAA^\prime B}$ from Eq.~\eqref{eq:rhoCAAB}, the final state takes the form 
\begin{widetext}
\begin{align}
    &\rho_{CAA^\prime B}^K \nonumber \\&= \sum_{\mu,\nu}M_{\mu\nu}^K\rho_{CAA^\prime B}(M_{\mu\nu}^K)^\dagger \nonumber \\
    &= |\alpha|^2 \ket{0}_C\langle0|\otimes \sum_\mu K_\mu \rho_{AA^\prime B}K_\mu^\dagger +  \frac{\alpha \beta^*}{4} \ket{0}_C\langle1|\otimes \sum_{\mu,\nu} K_\mu \rho_{AA^\prime B}K_\nu^\dagger  +\frac{\alpha^* \beta}{4} \ket{1}_C\langle0|\otimes \sum_{\mu,\nu} K_\nu \rho_{AA^\prime B}K_\mu^\dagger  \nonumber \\ &+|\beta|^2 \ket{1}_C\langle1|\otimes \sum_\nu K_\nu \rho_{AA^\prime B}K_\nu^\dagger .
    \label{kraus_on_state-app}
\end{align}
\end{widetext}
The first and last term of Eq.~\eqref{kraus_on_state-app} comes out to be 
\begin{align*}
    \sum_{\mu = 0}^3 K_\mu \rho_{AA^\prime B}K_\mu^\dagger = \psi^-_{AA^\prime}\otimes \left[ y\rho_d + (1-y)\sigma_x\rho_d\sigma_x\right]_B,
\end{align*}
where $y \coloneq p_1 + p_2$ , $\psi^- \coloneq |\psi^-\rangle\langle \psi^-|$ and $\rho_d \coloneq \frac{1}{2}(|\phi\rangle\langle\phi| + \sigma_z |\phi\rangle\langle\phi| \sigma_z)$. The other two terms of Eq.~\eqref{kraus_on_state-app} are as follows:  
\begin{align*}
    \sum_{\mu, \nu = 0}^3 K_\mu \rho_{AA^\prime B}K_\nu^\dagger =  \psi^-_{AA^\prime}\otimes \left[ 2y\sigma_z|\phi\rangle\langle\phi|\sigma_z\right]_B.
\end{align*} 
Now, applying Hadamard gate on control qubit $C$ and measuring it in the the computational basis, the state obtained at Bob is given by 
\begin{align}
    \tilde{\rho}^+_B &\coloneq \frac{y}{2}\rho_d + \frac{1-y}{2}\sigma_x\rho_d\sigma_x  + \frac{\alpha\beta^* + \alpha^*\beta}{4}y \sigma_z|\phi\rangle\langle\phi|\sigma_z \nonumber 
\\
    \tilde{\rho}^-_B &\coloneq  \frac{y}{2}\rho_d + \frac{1-y}{2}\sigma_x\rho_d\sigma_x  - \frac{\alpha\beta^* + \alpha^*\beta}{4}y \sigma_z|\phi\rangle\langle\phi|\sigma_z \nonumber
\end{align}
after partial tracing out the $CAA^\prime$ parts from $\rho_{CAA^\prime B}^K$. These states are not normalized. We further normalize these states. For brevity, let us set $X \coloneq \alpha\beta^* + \alpha^*\beta$. The normalized states obtained at Bob on clicking $\ket{0}$ and $\ket{1}$ at control system are  
\begin{align*}
    {\rho}^+_B &\coloneq \frac{2y\rho_d + 2(1-y)\sigma_x\rho_d\sigma_x  + Xy \sigma_z|\phi\rangle\langle\phi|\sigma_z}{2 + Xy} \hspace{3mm} \text{and}\nonumber \\
    {\rho}^-_B &\coloneq  \frac{2y\rho_d + 2(1-y)\sigma_x\rho_d\sigma_x  - Xy \sigma_z|\phi\rangle\langle\phi|\sigma_z}{2 - Xy} \hspace{3mm} \text{respectively}.
\end{align*}
The corresponding probabilities for clicking $\ket{0}$ and $\ket{1}$ are given by
\begin{align*}
    p_+ &= \frac{2 + Xy}{4} \quad \text{and} \quad 
    p_- =  \frac{2 - Xy}{4} \hspace{3mm} \text{respectively}.
\end{align*}
We can find the fidelity $\mathcal{F}_\pm^K \coloneq |\langle\phi|\rho^\pm_B|\phi\rangle|$ where
\begin{align*}
\mathcal{F}_\pm^K &= \frac{2y(1-2a^2)^2 + 4a^2(1-a^2) \pm Xy(1-2a^2)^2}{2\pm Xy}
\end{align*}
setting $a \coloneq \cos(n/2)$ with $0\leq n \leq \pi$ and integrating $\mathcal{F}^K_\pm$ over all pure states, we get 
\begin{align}
    \langle\mathcal{F}_\pm^K\rangle \coloneq\frac{1}{2} \int_0^\pi\mathcal{F}_\pm^K(n)\sin(n)dn = \frac{2y \pm Xy+2}{3(2\pm Xy)}.
    \label{eq-app-avg}
\end{align}
For quantum advantage in quantum teleportation, we need
\begin{align*}
    &\langle\mathcal{F}_\pm^K\rangle > \frac{2}{3} \\
    \Rightarrow &\frac{2y \pm Xy+2}{3(2\pm Xy)} - \frac{2}{3} > 0 \\
    \Rightarrow &y(2 \mp X) > 2.
\end{align*}
We show this region in blue color in Fig.~\ref{fig:Xy-schematic}.

\subsection{\texorpdfstring{Protocol characterized by Kraus operators $\{M_{\mu\nu}^L\}$}{Protocol characterized by Kraus operators M}}
\label{sec:Li}
Here, we examine the teleportation protocol assisted with the superposition of paths, characterized by the Kraus operators $\{M_{\mu\nu}^L\}$ defined in Eq.~\eqref{map-L}, when the shared state $\rho_{A^\prime B}$ is given by Eq.~\eqref{state1}.
The action of this protocol on the state of the joint control-Alice-Bob system is expressed by Eq.~\eqref{kraus_on_state-app}, with the only modification being the replacement of the operators $\{K_i\}$ by $\{L_i\}$, as defined in Eq.~\eqref{phi-kraus}.

Let us denote the state at Alice that she wants to teleport to Bob by $|\phi\rangle_A\langle\phi| $ and the state shared between Alice and Bob for teleportation by $\rho_{A^\prime B}$. $\rho_{AA^\prime B} (\coloneqq |\phi\rangle_A\langle\phi| \otimes \rho_{A^\prime B})$ represents the initial joint state of the Alice-Bob system. 
Note that 
\begin{align*}
    \sum_{\mu = 0}^3 L_\mu \rho_{AA^\prime B}L_\mu^\dagger &= \phi^-_{AA^\prime}\otimes \left[ z\rho_d + (1-z)\sigma_x\rho_d\sigma_x\right]_B, \\
     \sum_{\mu, \nu = 0}^3 L_\mu \rho_{AA^\prime B}L_\nu^\dagger &=  \phi^-_{AA^\prime}\otimes \left[ 2z\sigma_z|\phi\rangle\langle\phi|\sigma_z\right]_B,
\end{align*}
where 
$z \coloneq p_0 + p_3 = 1 -y$ , $\phi^- \coloneq |\phi^-\rangle\langle \phi^-|$ and $\rho_d = \frac{1}{2}(|\phi\rangle\langle\phi| + \sigma_z |\phi\rangle\langle\phi| \sigma_z)$.

Subsequently, the Hadamard gate is applied on the control qubit of the final joint state of the control-Alice-Bob system after this teleportation protocol, characterized by $\{M_{\mu\nu}^L\}$ followed by measuring the control qubit in the computational basis. The normalized states obtained at Bob on clicking $\ket{0}$ and $\ket{1}$ at control qubit are given by
\begin{align*}
    {\rho}^+_B &= \frac{2z\rho_d + 2(1-z)\sigma_x\rho_d\sigma_x  + Xz \sigma_z|\phi\rangle\langle\phi|\sigma_z}{2 + Xz} \hspace{2mm}\text{and}\nonumber \\
    {\rho}^-_B &=  \frac{2z\rho_d + 2(1-z)\sigma_x\rho_d\sigma_x  - Xz \sigma_z|\phi\rangle\langle\phi|\sigma_z}{2 - Xz}\hspace{2mm}\text{respectively.}
\end{align*}
The corresponding probabilities for getting ${\rho}^+_B$ and ${\rho}^-_B$ at Bob are
\begin{align*}
    p_+ = \frac{2 + Xz}{4},  \hspace{2mm}
    p_- =  \frac{2 - Xz}{4}\hspace{2mm}\text{respectively}.
\end{align*}
Therefore, integrating over all pure single-qubit states at Alice, the average fidelity will have the following form:
\begin{align*}
    \langle\mathcal{F}_\pm^L\rangle  \coloneqq \frac{2z \pm Xz+2}{3(2\pm Xz)}.
\end{align*}
By substituting $z = 1 - y$, we obtain that the condition $(1 - y)(2 \mp X) > 2$ must be satisfied to achieve quantum advantage for the class of separable shared states given in Eq.~\eqref{state1}, using our protocol defined by the Kraus operators $\{M_{\mu\nu}^L\}$. The region corresponding to this quantum advantage is highlighted in green color in Fig.~\ref{fig:Xy-schematic}.

% {\color{black}
% \section{Equivalence between Eq.~\eqref{typegen} and Eqs.~\eqref{type2}~and~\eqref{type1} via local unitary}
% ~\textcolor{red}{[Plz write here what you want to write here. paper er vitore ki mention kora achhe j ai kotha ta appendix e achhe?]}
% % \begin{align*}
% %     \rho_{A^\prime B} = r |\omega \gamma\rangle\langle \omega \gamma| + (1-r) |\bar{\omega} \bar{\gamma}\rangle \langle \bar{\omega} \bar{\gamma}|,
% % \end{align*}

% % \begin{align*}
% %     \rho_{A^\prime B} = p |00\rangle\langle 00| + (1-p) |11\rangle \langle 11|,
% % \end{align*}
% %  \begin{align*}
% %     \rho_{A^\prime B} = q |01\rangle\langle 01| + (1-q) |10\rangle \langle 10|,
% % \end{align*}
% }

\section{\texorpdfstring{Protocol for implementing standard teleportation with superposition of paths for separable states of the form in Eq.~\eqref{typegen}}{Teleportation protocol with superposition of paths}}
\label{sec:Advantage pure product}
Here, we investigate how the teleportation protocol can be implemented in conjunction with superposition of paths  for separable states of the form of Eq.~\eqref{typegen}. As discussed in previous section we want to teleport a qubit pure state $|\phi\rangle$. 
The shared state is of the form 
\begin{align}
    \rho_{A^\prime B} = r|\omega \gamma \rangle \langle \omega\gamma| + (1-r) |\bomga\rangle\langle\bomga|,
\end{align}
Both the subsystems $A^\prime$ and $B$ are described by states in two-dimensional Hilbert spaces.  The states $\{|\omega\rangle, \ket{\bom}\}$ and $\{\ket{\gamma}, \ket{\bga} \}$ form orthonormal basis states in their respective Hilbert spaces.  
Our protocol is realized through the Kraus operators of the following form 
\begin{align*}
    E_{\mu\nu} = \ket{0}_C \langle 0| \otimes \tilde{K}_\mu + \ket{1}_C \langle 1| \otimes \tilde{K}_\nu. 
\end{align*}
Where the Kraus operators $\{\tilde{K}_\mu\}$ are for the standard teleportation protocol and they act on the parties $AA^\prime B$. Their explicit forms are as follows
\begin{align}
\label{gen-kraus}
    \tilde{K}_0 &= (|\beta_0\rangle\langle\beta_0|)_{AA^\prime}\otimes (\mathbb{I}_2)_B, \nonumber \\
    \tilde{K}_1 &= (|\beta_0\rangle\langle\beta_1|)_{AA^\prime}\otimes (U_1)_B, \nonumber \\
    \tilde{K}_2 &= (|\beta_0\rangle\langle\beta_2|)_{AA^\prime}\otimes (U_2)_B ,\nonumber \\
    \tilde{K}_3 &= (|\beta_0\rangle\langle\beta_3|)_{AA^\prime}\otimes (U_3)_B,
\end{align}
where, $|\beta_0\rangle = \frac{1}{\sqrt 2} (|0\omega\rangle - |1\bom\rangle)$, $|\beta_1\rangle = \frac{1}{\sqrt 2} (|0\omega\rangle + |1\bom\rangle)$, $|\beta_2\rangle = \frac{1}{\sqrt 2} (|0\bom\rangle - |1\omega\rangle)$ and $|\beta_3\rangle = \frac{1}{\sqrt 2} (|0\bom\rangle + |1\omega\rangle)$ with $U_1|\gamma\rangle = |\gamma\rangle$, $U_1|\bga\rangle = -|\bga\rangle$, $U_2|\gamma\rangle = |\bga\rangle$, $U_2|\bga\rangle = |\gamma\rangle$ and $U_3|\gamma\rangle = -|\bga\rangle$, $U_3|\bga\rangle = |\gamma\rangle$.

The joint initial state of control-Alice-Bob is given by $\rho_{CA A^\prime B} 
 \coloneq \chi_c \otimes \phi_{A} \otimes \rho_{A^\prime B}$, with $\phi \coloneq |\phi\rangle \langle \phi|$ and $\chi \coloneq |\chi\rangle \langle \chi |$. The state $|\phi\rangle$ is to be teleported form Alice to Bob, while the state $|\chi\rangle$ is the initial control qubit state. 
After operating the Kraus operators  $\{E_{\mu\nu}\}$, we apply the Hadamard gate on the control qubit and measure it in the computational basis. The normalized states obtained at Bob on getting $\ket{0}$ and $\ket{1}$ in the control qubit are given by
\begin{align*}
    {\rho}^+_B &= \frac{2\tilde\rho_d +  X \sigma_z|\tilde\phi\rangle\langle\tilde\phi|\sigma_z}{2 + X} \hspace{2mm} \text{and}\nonumber \\
    {\rho}^-_B &=  \frac{2\tilde\rho_d  - X \sigma_z|\tilde\phi\rangle\langle\tilde\phi|\sigma_z}{2 - X}\hspace{2mm} \text{respectively.}
\end{align*}
Here $\tilde\rho_d = \frac{1}{2}(|\tilde\phi\rangle\langle\tilde\phi| + \sigma_z |\tilde\phi\rangle\langle\tilde\phi| \sigma_z)$ and $|\tilde\phi\rangle = U_\gamma|\phi\rangle$, with $U_\gamma\ket{0} = |\gamma\rangle$ and $U_\gamma\ket{1} = |\bga\rangle$.
The corresponding probabilities are of obtaining $|0\rangle$ and $|1\rangle$ at the control are 
\begin{align*}
    p_+ &= \frac{2 + X}{4} \quad \text{and} \quad
    p_- =  \frac{2 - X}{4}\hspace{2mm} \text{respectively}.
\end{align*}
Note that when $X = \mp 1$, state obtained at Bob is $\rho_B^\pm = |\tilde \phi \rangle$ with probability $p_\pm = \frac{1}{4}$. This is not exactly the state $|\phi\rangle$ that was supposed to be teleported; rather, this obatined state is unitarily rotated from $|\phi\rangle$ with the unitary $U_\gamma$. Hence, at the end of the protocol, Bob has to apply locally the $U_\gamma^\dagger$ operator on the state $|\tilde\phi\rangle$ that is obtained. This is not difficult, as Bob has the knowledge of his initial state which was prepared as mixture of basis states $\{|\gamma\rangle, |\bga\rangle\}$.    

{\color{black}
\section{Teleportation with superposition of indefinite causal order for shared pure product states}
\label{teleportation-switch-app}
Here we apply the teleportation in conjunction with  superposition of indefinite causal order. 
We first use the theorem in Ref.~\cite{Bowen2001}, which states that the standard teleportation protocol is equivalent to a generalized depolarizing channel $\Lambda$, that is, if we want to teleport the state $\rho$, the output state under the action of the standard teleportation protocol is given by $\Lambda(\rho) = \sum_{i = 0}^3 p_i \sigma_i \rho\sigma_i$. 
We have  $ p_i = \langle \beta_i| \gamma_{\text{shared}} |\beta_i\rangle$,
where $\gamma_{\text{shared}}$ denotes the shared state using which we want to teleport state from Alice to Bob. $|\beta_0\rangle$ denotes the maximally entangled state having largest overlap with the shared state $\gamma_{\text{shared}}$ (that is $p_0$ is maximum) and all other states $|\beta_i\rangle$ for $i \neq 0$ are obtained by applying the Pauli operators locally on one of the parties of the state $|\beta_0\rangle$. 

Let us consider a pure product state of the form of Eq.~\eqref{typegen} with $r = 1$, that is the shared state is $\rho_{A^\prime B} = |\omega\gamma\rangle\langle \omega \gamma|$. 
For such a state the singlet fraction $p_0 = \langle \beta_0 | \rho_{A^\prime B} |\beta_0\rangle = \frac{1}{2}$ and $p_3 = \langle \beta_3 | \rho_{A^\prime B} |\beta_3\rangle = \frac{1}{2}$, while $\langle \beta_1|\rho_{A^\prime B}|\beta_1\rangle = \langle \beta_2|\rho_{A^\prime B}|\beta_2\rangle = 0$. Here, $\beta_0 = \frac{1}{\sqrt 2} |\omega\gamma\rangle + |\bar\omega \bar\gamma\rangle$, $\beta_1 = \frac{1}{\sqrt 2} |\omega\bar\gamma\rangle + |\bar\omega \gamma\rangle$, $\beta_2 = \frac{1}{\sqrt 2} |\omega\bar\gamma\rangle - |\bar\omega \gamma\rangle$, and $\beta_3 = \frac{1}{\sqrt 2} |\omega\gamma\rangle - |\bar\omega \bar\gamma\rangle$.
Thus, the generalized depolarizing map corresponding to the teleportation protocol is given by
\begin{align*}
    \mathcal{E}(\rho) = p_0 \rho + p_3\sigma_z\rho\sigma_z,
\end{align*}
with $p_0 = p_3 = \frac{1}{2}$ when such a pure product state is being shared between Alice and Bob. 

Now, the Kraus operators for the operation of superposition of indefinite causal order of $\mathcal{E}$ with itself will have the following form: $S_{ij} = K_iK_j\otimes \ket{0}_C\langle0| + K_jK_i\otimes \ket{1}_C\langle1|$, with $\sum_{i,j} S_{ij}^\dagger S_{i,j} = \mathbb{I}_4$. Here, $\{\ket{0}, \ket{1}\}$ represents a basis in control's Hilbert space. In our case, $K_0 = \frac{1}{\sqrt 2} \mathbb{I}_2$ and $K_1 = \frac{1}{\sqrt 2} \sigma_z$. Let $|\chi\rangle_C \coloneqq \alpha\ket{0} + \beta\ket{1}$ be the initial control qubit state, and $|\phi\rangle_S$ is the state that we want to teleport. Therefore, the initial state of the control-target system will be $|\chi\rangle_C \otimes |\phi\rangle_S$.

Then, the final joint state of the control-target system under the action of our protocol characterized by $S_{ij}$ on the initial state of the control-target system will be
\begin{align*}
    \rho_{SC}^\prime =  \sum_{i,j = 0}^1 S_{i,j} |\phi\rangle_S\langle\phi|\otimes |\chi\rangle_C\langle\chi|S_{ij}^\dagger, 
\end{align*}
From this we obtain
\begin{align*}
    \rho_{SC}^\prime = \rho_d \otimes |\chi\rangle_C\langle\chi|.
\end{align*}
Therefore, the final teleported state turns out to be $\rho_S' = \rho_d \coloneqq\frac{1}{2}(|\phi\rangle\langle\phi| + \sigma_z |\phi\rangle\langle\phi|\sigma_z)$.

The fidelity with the initial state $\ket{\phi}$ is
$\tilde{\mathcal{F}} = \langle\phi|\rho_S'|\phi\rangle = \frac{1}{2}\left(1 + (2a^2-1)^2\right),$
where we have $|\phi\rangle = a|0\rangle +e^{i\theta}\sqrt{1-a^2}|1\rangle$.
Following Eq.~\eqref{eq-app-avg} and setting $a = \cos(n/2)$, the average fidelity turns out to be
$$\langle \tilde{\mathcal{F}}\rangle = \frac{1}{2} \int_0^\pi\tilde{\mathcal{F}}(n)\sin(n)dn = \frac{2}{3}.$$
}

{\color{black}
\section{Standard teleportation protocol with shared state of the form of Eq.~\eqref{typegen}}
\label{standard-tele-shared-sep}
We here show that the shared state of the form of Eq.~\eqref{typegen}, when used in standard teleportation protocol provides the average fidelity of $\frac{2}{3}$, and hence acts as at the optimal limit classically achievable. 
We consider the shared state of Eq.~\eqref{typegen}, as follows
$$ \rho_{A^\prime B} = r|\omega \gamma \rangle \langle \omega\gamma| + (1-r) |\bomga\rangle\langle\bomga|.$$ 
To implement the standard teleportation protocol, we apply the Kraus operators $\{\tilde{K}_\mu \}$ as presented in Eq.~\eqref{gen-kraus}. As usual, the state that is to be teleported is given by $|\phi\rangle$. Thus, we get
\begin{align*}
    \rho_{AA'B}' &= \sum_\mu \tilde K_\mu (|\phi\rangle\langle\phi|\otimes \rho_{A'B}) \tilde K_\mu^\dagger\\
    &= |\beta_0\rangle\langle \beta_0|\otimes \tilde{\rho}_d.
\end{align*}

Here, $\tilde\rho_d = \frac{1}{2}(|\tilde\phi\rangle\langle\tilde\phi| + \sigma_z |\tilde\phi\rangle\langle\tilde\phi| \sigma_z)$ and $|\tilde\phi\rangle = U_\gamma|\phi\rangle$, with $U_\gamma\ket{0} = |\gamma\rangle$ and $U_\gamma\ket{1} = |\bga\rangle$. Thus applying $U_\gamma^\dagger$ on $\tilde\rho_d$, we finally get the teleported state to be $\rho = \frac{1}{2}(|\phi\rangle\langle\phi| + \sigma_z |\phi\rangle\langle\phi| \sigma_z)$. 
The fidelity with the initial state $\ket{\phi}$ is
${\mathcal{F}} = \langle\phi|\rho_S|\phi\rangle = \frac{1}{2}\left(1 + (2a^2-1)^2\right),$
where we have $|\phi\rangle = a|0\rangle +e^{i\theta}\sqrt{1-a^2}|1\rangle$.
Following Eq.~\eqref{eq-app-avg} and setting $a = \cos(n/2)$, the average fidelity turns out to be
$$\langle {\mathcal{F}}\rangle = \frac{1}{2} \int_0^\pi{\mathcal{F}}(n)\sin(n)dn = \frac{2}{3}.$$
}

{\color{black}
\section{Generalization to higher dimensions}
\label{higer-dim-generalizations}
Here we generalize our scheme of teleporting a qudit in conjunction with superposition of paths. A qudit state $|\psi\rangle = \sum_{i = 0}^{d-1}\alpha_i |i\rangle$, is to be teleported. 

First let us briefly mention the Kraus operators of teleporting a qudit in the conventional teleportation protocol. The party $A$ has the state $|\psi\rangle_A$ and the parties $A'B$ share the maximally entangled state of $|\phi_{0k}\rangle_{A'B}$. Here, $|\phi_{0k}\rangle = \frac{1}{\sqrt d} \sum_{j = 0}^{d-1} \omega^{kj}|jj\rangle$, where $\omega = e^{2\pi i/d}$ is the $d$-th root of unity. 
The Kraus operator for teleporting the state $|\psi\rangle_A$ from $A$ to $B$ is given by,
\begin{align}
    \label{Karus-qudit}
    K^{(k)}_{mn} = |\phi_{0k}\rangle_{AA'}\langle\phi_{mn}| \otimes (Z^{-k}Z^{n} X^{-m})_B.
\end{align}
Here, $|\phi_{mn}\rangle = \frac{1}{\sqrt d}\sum_{j = 0}^{d-1} \omega^{jn}|j\rangle|j+m\rangle$, $j+m$ is in mod$(d)$. The Weyl operators act as, $X^m|j+m\rangle$ and $Z^n|j\rangle = \omega^{jn}|j\rangle$. In the conventional teleportation protocol, the initial state is
\begin{align}
    |\psi\rangle_A|\phi_{0k}\rangle_{A'B} = \frac{1}{d}\sum_{ab}|\phi_{ab}\rangle_{AA'} \otimes Z^k X^a Z^{-b}|\psi\rangle_B.
\end{align}
Applying the Kraus operator on the initial state, we get,
\begin{align*}
    K^{(k)}_{mn} |\psi\rangle_A &|\phi_{0k}\rangle_{A'B}  = \frac{1}{d}\omega^{mk}|\phi_{0k}\rangle_{AA'} |\psi\rangle_B.
\end{align*}
Thus, we get 
\begin{align*}
    \sum_{mn}K^{(k)}_{mn} \sigma_{AA'B}{K_{mn}^{(k)}}^\dagger = |\phi_{0k}\rangle_{AA'}\langle\phi_{0k}|\otimes |\psi\rangle_B\langle\psi|
\end{align*} 
Here, $\sigma_{AA'B} = |\psi\rangle_A\langle\psi|\otimes|\phi_{0k}\rangle_{A'B}\langle\phi_{0k}|$. Thus, the state got teleported from $A$ to $B$.
% Here, we have used $ZX = \omega XZ$ and $\sum_{mn}\omega^{mn} = d$.

Now, let us consider the teleportation of qudits in conjunction with superposition of paths. We consider the initial state of the parties $AA'B$ as $|\Psi\rangle_{AA'B} = |\psi\rangle_A|00\rangle_{A'B}$ such that $\rho_{AA'B} = |\Psi\rangle_{AA'B}\langle\Psi|$. In analogy to the qubit case, we construct the Kraus operators as
\begin{align}
    M_{mnab} = |0\rangle_C\langle0|\otimes\frac{1}{d}K_{mn}^{(k)} + |1\rangle_C\langle1|\otimes\frac{1}{d}K_{ab}^{(k)}
\end{align}
such that $\sum_{mnab}M_{mnab}^\dagger M_{mnab} = \mathbb{I}_{2d^3}$. Following Eq.~\eqref{eq:total_operation}, we need to evaluate $\sum_{mn}K_{mn}^{(k)}\rho_{AA'B}{K_{mn}^{(k)}}^\dagger$ and $\sum_{mnab}K_{mn}^{(k)}\rho_{AA'B}{K_{ab}^{(k)}}^\dagger$. We note that,
\begin{align*}
    |\Psi\rangle_{AA'B} &= \frac{1}{d^{3/2}}\sum_{abc}|\phi_{ab}\rangle_{AA'}\otimes Z^cX^aZ^{-b}|\psi\rangle_B,\\
K_{mn}^{(k)}|\Psi\rangle_{AA'B} &= \frac{1}{d^{3/2}}|\phi_{0k}\rangle_{AA'}\otimes\sum_{c}Z^{-k}Z^c \omega^{-mc}|\psi\rangle_B \\
&= \frac{1}{\sqrt d}|\phi_{0k}\rangle_{AA'}\otimes Z^{-k} \alpha_{m}|m\rangle_B.
\end{align*}
Here we haves used the following,
\begin{align*}
    \sum_c Z^c \omega^{-mc}|\psi\rangle &= \sum_{c,i}\omega^{c(i-m)}\alpha_i|i\rangle\\
    &= \sum_i d\delta_{i,m}\alpha_i|i\rangle = d\alpha_m|m\rangle.
\end{align*}
Thus, we get 
\begin{align*}
  \sum_{mn}K_{mn}^{(k)} |\Psi\rangle_{AA'B}\langle\Psi|{K_{mn}^{(k)}}^\dagger &=  |\phi_{0k}\rangle_{AA'}\langle\phi_{0k}|\otimes\tilde{\rho_d}_B \\
\sum_{mnab}K_{mn}^{(k)} |\Psi\rangle_{AA'B}\langle\Psi|{K_{ab}^{(k)}}^\dagger &= \left[\sum_{mn}K_{mn}^{(k)} |\Psi\rangle\right]\left[\sum_{ab}\langle\Psi|{K_{ab}^{(k)}}^\dagger\right]\\
&= {d}|\phi_{0k}\rangle_{}\langle\phi_{0k}|\otimes Z^{-k}|\psi\rangle_{}\langle\psi|Z^k.
\end{align*}
Here, ${\tilde\rho_{d}} = \sum_{i}|\alpha_i|^2|i\rangle\langle i|$, is the diagonal part of the state to be teleported. Thus, after applying the Kraus operators in analogy to Eq.~\eqref{eq:total_operation}, and then measuring the control qubit $C$ in $\{|+\rangle,|-\rangle\}$ basis, we get 
\begin{align*}
    \rho^\pm &\coloneqq \frac{d\tilde\rho_d \pm \mathcal{B}Z^{-k}|\psi\rangle\langle\psi|Z^k}{d \pm\mathcal{B}},\\
    p^\pm &\coloneqq \frac{d \pm \mathcal{B}}{2d}.
\end{align*}
Here, $\mathcal{B}\coloneqq \alpha\beta^* + \beta\alpha^*$.
Setting $d = 2$, we get back the qubit case, discussed in rest of the paper. The average fidelity turns out to be 
\begin{align*}
    \langle\mathcal{F}^{\pm}_d\rangle &=  \frac{d}{d\pm\mathcal B} \Big\langle\langle\psi|\tilde\rho_d|\psi\rangle\Big \rangle_{\text{Pure}} \pm \frac{\mathcal B}{d\pm\mathcal B} \Big\langle |\langle \psi| Z^k|\psi\rangle|^2 \Big\rangle_{\text{Pure}}  \\
    &= \frac{2d \pm \mathcal B}{(d+1)(d\pm \mathcal B)}.
\end{align*}
We have used the fact that $\tilde\rho_d = d^{-1}\sum_n Z^n|\psi\rangle\langle\psi|Z^{-n}$ and 
\begin{align*}
\Big\langle |\langle \psi | Z^a | \psi \rangle|^2\!\Big\rangle_{\text{Pure}}
=
\begin{cases}
\dfrac{1}{d+1}, & a \neq 0, \\
1, & a = 0.
\end{cases}
\end{align*}
$\Big\langle\cdot\Big\rangle_\text{Pure}$ is the average over all $d$-dimensional pure states.
We choose our initial control qubit state such that $\mathcal B = -1$ and choose that the $|+\rangle$ state after measurement in $C$. Then we get,
$$\langle\mathcal{F}^{\pm}_d\rangle = \frac{2d-1}{d^2 - 1} > \mathcal F_{cl}.$$
The optimal fidelity obtained using classical resources, $\mathcal F_{cl} = 2/(d+1)$~\cite{horodecki1999}. Thus, we can achieve average fidelity better than the optimal classical protocol. 
}
{\color{black}
\section{Average fidelity bound for classical strategies with post-selection}
\label{final-app}
Here, we determine the average fidelity achieved when a qubit is teleported using only classical resources, with post-selection allowed. We show that post-selection does not enhance the average fidelity of such classical strategies.
 % We first clarify a subtle point. One might argue that by restricting attention to states in a small neighborhood of $\{\ket{0},\ket{1}\}$ and rejecting all other inputs, the conditional (post-selected) fidelity can be made arbitrarily close to $1$ as the success probability tends to zero. In the limiting case, only the states $\{\ket{0},\ket{1}\}$ are accepted, while all others are discarded. Then, if Alice measures in the computational basis, Bob can reproduce the state with unit fidelity, albeit with vanishing success probability.

% We do not consider such a scenario. In the standard teleportation setting, Alice is given only a single copy of $\ket{\phi}$, and therefore cannot determine whether the state lies in (or is close to) a particular region of the Bloch sphere, such as the neighborhood of the computational basis states. Consequently, such selective post-processing strategies are not operationally meaningful in this context. Instead, we restrict our analysis to general measure-and-prepare strategies undertaken by Alice and Bob.

 A general classical strategy for transmitting $\ket{\phi}$ from Alice to Bob proceeds as follows. Alice performs a general POVM $\{M_i\}$ on the given state and communicates the outcome $i$ to Bob. Unlike the standard teleportation protocol, where only two classical bits are sent, here Alice is allowed to transmit an arbitrary amount of classical information, thereby encoding $i$ without restriction. 

Upon receiving the message, Bob prepares a state $\sigma_i$ conditioned on the outcome $i$. Additionally, Alice is allowed to post-select her outcomes; that is, she communicates with Bob only if the measurement outcome belongs to a predefined success set $S$, i.e., $i \in S$. 

Accordingly, the average output state prepared by Bob is given by
\begin{equation}
\rho_B^{\text{out}} = \frac{\sum_{i \in S} \mathrm{Tr}(M_i \phi)\,\sigma_i}{\sum_{i \in S} \mathrm{Tr}(M_i \phi)}\,,
\end{equation}
where $\phi = \ket{\phi}\bra{\phi}$. The success probability is $p(\phi, S) = \sum_{i \in S}\text{Tr}[M_i\phi]$.  We define the conditional fidelity of this state as
\begin{align}
     \mathcal{F}_S^\phi &= \langle\phi|\rho_B^\text{out}|\phi\rangle = \frac{\sum_{i\in S}\text{Tr}(M_i \phi)\langle\phi|\sigma_i|\phi\rangle}{\sum_{i\in S} \text{Tr} (M_i \phi)} \nonumber \\
     &= \frac{\sum_{i\in S}\text{Tr}[(M_i\otimes\sigma_i)(\phi\otimes \phi)]}{\sum_{i\in S} \text{Tr} (M_i \phi)}.
 \end{align}
Now, we average $\mathcal{F}^\phi_S$ over all states using the Haar-measure. We note that since post-selection conditions on the event of success, the ensemble of input states associated with successful runs is no longer uniform, but is instead described by the conditional distribution, $p(\phi|S) = p(\phi,S)/p(S)$. Now, we know that $p(\phi,S) = \sum_{i \in S} \text{Tr}(M_i \phi)$ and $p(S) = \int d\phi \sum_{i \in S} \text{Tr}(M_i \phi)$. The average fidelity turns out to be,
\begin{align}
    \langle \mathcal{F}_S\rangle = \int d\phi \mathcal{F}_S^\phi p(\phi|S) = \frac{\int d\phi \sum_{i\in S}\text{Tr}[(M_i\otimes\sigma_i)(\phi\otimes \phi)]}{\int d\phi \sum_{i \in S} \text{Tr}(M_i \phi)}
\end{align}
Now we do the Haar-average as $\int d\phi \ \phi = {\mathbb{I}}/{2}$ and $\int d\phi \ (\phi\otimes\phi) = {(\mathbb{I}+\mathbb{S}) }/{6}$~\cite{Mele2024introductiontohaar}, where $\mathbb{S}$ is the swap(flip) operator. Using the fact $\text{Tr}[\mathbb{S}(A\otimes B)] = \text{Tr}[AB]$, we get
\begin{align}
    \langle \mathcal{F}_S\rangle &= \frac{\frac{1}{6}(\sum_{i\in S}\text{Tr}[M_i\sigma_i] + \text{Tr}[M_i])}{\frac{1}{2}(\sum_{i\in S}\text{Tr}[M_i])}\nonumber\\
    &\leq \frac{\frac{1}{6}(\sum_{i\in S} 2\text{Tr}[M_i])}{\frac{1}{2}(\sum_{i\in S}\text{Tr}[M_i])} = \frac{2}{3}.
\end{align}
Here we use the fact that $\mathbb{O}\leq\sigma_i \leq \mathbb{I}$, so $\text{Tr}[M_i\sigma_i]\leq\text{Tr}[M_i]$.
Thus, we see that $\langle \mathcal{F}_S\rangle \leq 2/3$, hence post-selection with only classical resources do not provide any advantage in teleportation tasks.
}
\bibliography{References}

\end{document}